\documentclass[trackchanges]{aastex701}

\usepackage{orcidlink}

\begin{document}

\title{Aditya-L1 and SDO Observations of a Confined X-Class Flare with an EUV Late Phase: Thermal and Nonthermal Emission, Magnetic Topology, and Flux Rope Confinement}

\correspondingauthor{Vishwa Vijay Singh}
\email{vishwavijaysingh630@gmail.com}

\author[0009-0005-3692-1893]{Vishwa Vijay Singh}
\affiliation{Udaipur Solar Observatory, Physical Research Laboratory, Udaipur 313 001, Rajasthan, India}
\affiliation{Discipline of Physics, Indian Institute of Technology Gandhinagar, Palaj, Gandhinagar-382 355, Gujarat, India}
\email{vishwavijaysingh630@gmail.com}

\shortauthors{Singh et al.}
%\author{Bhuwan Joshi\,\orcidlink{0000-0001-5042-2170}}
\author[0000-0001-5042-2170]{Bhuwan Joshi}
\affiliation{Udaipur Solar Observatory, Physical Research Laboratory, Udaipur 313 001, Rajasthan, India}
\email{bhuwan@prl.res.in}

%\author{N.P.S. Mithun\,\orcidlink{0000-0003-3431-6110}}
\author[0000-0003-3431-6110]{N.P.S. Mithun}
\affiliation{Physical Research Laboratory, Navrangpura, Ahmedabad, 380 009, India}
\email{mithun@prl.res.in}

%\collaboration{all}{The Terra Mater collaboration}

%% Use the \collaboration command to identify collaborations. This command
%% takes an optional argument that is either a number or the word "all"
%% which tells the compiler how many of the authors above the command to
%% show. For example "\collaboration[all]{(DELVE Collaboration)}" wil include
%% all the authors above this command.
%%
%% Mark off the abstract in the ``abstract'' environment. 
\begin{abstract}

We study a long-duration, confined X2.0-class solar flare exhibiting three distinct episodes of HXR emission, along with an accompanying EUV late phase. A key highlight of this work is the novel observations from the high-resolution solar X-ray spectrometers SoLEXS and HEL1OS onboard Aditya-L1, providing insights into the evolution of thermal and non-thermal emissions. Simultaneous EUV observations from AIA/SDO in hot channels, combined with multi-band X-ray light curves, reveal a two-stage confined eruption of a large MFR, separated by $\sim$10~minutes. The first stage corresponds to an M4.6 flare, while the overall event reaches X2.0 on the GOES scale. During the first phase, magnetic reconnection beneath the large EUV hot channel causes plasma heating to $\sim$19~MK, with a non-thermal spectral index of $\sim$4.3. In the second phase, characterized by external reconnection and confinement of the MFR hot channel, the temperature increases to $\sim$22~MK, accompanied by further hardening of the spectral index to $\sim$3.9. Despite pronounced non-thermal signatures and enhanced particle acceleration, the MFR fails to erupt through the overlying coronal loops because of the strong strapping force exerted by the large overlying coronal loops. Importantly, the stalled eruption, together with reconnection in the higher overlying loops, results in enhanced heating within a closed magnetic configuration, with temperatures peaking at up to $\sim$15~MK during the late phase. The paper also provides the first observations of detectable non-thermal X-ray emission during the early EUV late phase, albeit briefly, providing direct evidence for reconnection-induced heating, likely occurring in the higher overlying coronal loops.

\end{abstract}

%%
%% You can use the \uat command to link your UAT concepts back its source.

\section{Introduction}
\label{sec:Introduction}
Solar flares are the rapid release of magnetic energy in the solar atmosphere, converting stored magnetic free energy into plasma heating, bulk motions, and the acceleration of particles to suprathermal and relativistic energies. Large flares typically release on the order of $10^{32}$--$10^{33}$~erg over timescales of seconds to many minutes, producing bright emission across the electromagnetic spectrum from radio to $\gamma$-rays \citep[e.g.][]{Fletcher2011, Benz2017, joshi2024}. The underlying physical driver is magnetic reconnection, which reorganises coronal magnetic connectivity and facilitates the rapid conversion of magnetic energy into thermal and kinetic energy, as well as into accelerated particles \citep{priestandforbes2002, Su2013}.

While many major solar flares are accompanied by Coronal Mass Ejections \citep[CMEs; see reviews by][]{Forbes2000, Hudson2006, Chen2011} classified as eruptive flares, some flares do not lead to CMEs and are therefore classified as confined flares. A notable subset of confined events consists of so-called failed flux-rope eruptions, in which a magnetic flux rope (MFR) is initially activated and begins to rise but subsequently fails to escape \citep{Ji2003, Alexander2006, Gilbert2007, Liu2009, Kushwaha2015}. The successful eruption of an MFR from the Sun into the interplanetary medium is a key requirement for the formation of a CME. Broadly, two classes of models describe how an MFR transitions from a stable to an eruptive state: those based on ideal MHD instabilities and those relying on magnetic reconnection \citep[see, e.g.,][]{priestandforbes2002, Aulanier2014, Green2018}. In the ideal-instability scenario, a pre-existing MFR can erupt when the overlying magnetic field decreases sufficiently rapidly with height, leading to torus instability \citep{Kliem2006}, or when the twist of the flux rope exceeds a critical threshold, giving rise to kink instability \citep{Torok2004}. Reconnection-driven models, on the other hand, consider different pre-flare magnetic configurations of the active region (AR). The tether-cutting model assumes a bipolar topology in which the initial reconnection occurs within the highly sheared core fields, thereby triggering the eruption \citep{Moore1992, Moore2001}. By contrast, the magnetic breakout model requires a multi-polar environment with one or more coronal null points \citep{Karpen2012}, where reconnection above the core fields removes the confining tension of the overlying arcade \citep{Antiochos1999}. Building on these frameworks, early dynamics and configuration of MFR together with the ambient magnetic field can provide a possible explanation for failed eruption \citep[e.g.][]{Mitra2022}. 
 
Complementary evidence from data-driven MHD simulations by \citet{Zhong2021} shows that non-axisymmetric Lorentz-force components associated with the MFR geometry can substantially hinder its upward motion, offering a physical mechanism for failed eruptions even when classical instability thresholds are satisfied. The outcome of a filament eruption---full, partial, or failed---is strongly influenced by kinking motions and the location of magnetic reconnection. \citet{Gilbert2001, Gilbert2007} found that whether an eruption succeeds depends critically on where reconnection occurs relative to the prominence: reconnection below the flux rope promotes full eruptions; reconnection above it splits the flux rope and prevents mass escape (failed eruption); and reconnection within the prominence yields partial eruptions. Both partial and failed eruptions frequently exhibit filament re-formation as material falls back after the event. Many studies have examined the confined X3.1 flare from AR~12192, which was the most energetic event in the flare sequence of this active region. \citet{Sun2015} and \citet{Sarkar2018} analysed the magnetic configuration and reported that the core of AR~12192 exhibited relatively low non-potentiality with only minor flare-related variations, and that the confinement of the flare can be attributed to the presence of a strong overlying magnetic field. More recently, \citet{Vasantharaju2023} suggested that although the sigmoid--filament system underwent an upward catastrophic rise driven by mass unloading, the eruption was ultimately suppressed by the strong confining external poloidal field.

In addition to these instability-related factors, magnetic reconnection between the rising filament and the overlying coronal field has also been proposed as a mechanism that can inhibit or halt the eruption \citep{Amari1999,Hassanin2016}. Together, these results highlight that multiple physical processes—including instabilities, MFR geometry, Lorentz-force balance, weak non-potentiality,  and reconnection—jointly determine whether an eruption succeeds or is ultimately confined. Although energetic flares are often accompanied by large-scale coronal eruptions (i.e., CMEs), not all intense or long-duration flares generate CMEs. Observationally, confined flares typically exhibit an impulsive soft X-ray (SXR) light curve, implying that plasma cooling dominates following the initial energy release. In contrast, eruptive flares often show a prolonged, gradual decay phase lasting several hours, commonly referred to as long-duration events (LDEs; \citealt{Sheeley1983,Webb1987}), indicating sustained energy release well beyond the impulsive phase. We note that the long-duration events (LDEs) are typically linked to sustained reconnection in extended current sheets above post-flare arcades and therefore frequently produce CMEs and supra-arcade downflows \citep{McKenzie2000,yashiro2006}. High-resolution EUV imaging from SDO/AIA has revealed hot channel–like flux-rope structures prior to and during eruptions, but the presence of such a hot channel (a magnetic flux rope candidate) does not guarantee CME formation \citep{aia,Zhang2012,cheng2013}. Observational surveys show that while flare–CME coupling is common, the association is not universal: only about 75\% of GOES flares~$\ge$~X1.0 produce CMEs, increasing to $>$90\% for flares~$\ge$~X2.5 \citep{yashiro2006}.

X-ray observations provide some of the most direct probes of thermal and non-thermal flare processes. Thermal X-ray and EUV emission trace hot plasma generated by impulsive heating and chromospheric evaporation, while hard X-ray (HXR) bremsstrahlung offers a relatively unambiguous measure of accelerated electron populations \citep{Holman2011}. In the classical thick-target picture, the HXR photon spectrum reflects the properties of electrons that lose their energy collisionally in the dense lower atmosphere \citep{Brown1971}. Observations have revealed HXR emission can be directly associated with magnetic flux ropes, including emission up to 25~keV from an activated MFR \citep{Suraj2020} and intense non-thermal emission up to $\sim$200~keV with a hard spectral index ($\gamma \approx 3$) even in confined events \citep{Kushwaha2014}. Seminal Reuven Ramaty High Energy Solar Spectroscopic Imager (RHESSI) studies provided major insights into HXR emission and particle acceleration \citep{Fletcher2011,Holman2011,Kontar2011}. Although this framework is well established, a comprehensive quantitative understanding of the flare energy budget---particularly the relation between non-thermal electrons and early flux-rope evolution is poorly understood. 

Most of the earlier studies on the solar flare X-ray emission, based largely on single instruments data, often unable to simultaneously constrain both the hot ($>10$~MK) thermal component seen in HXR and the cooler ($\sim$2--10~MK) component dominating the SXR emission, the latter containing a substantial but often under-quantified fraction of the total energy \citep{Warmuth2020}. A more complete picture requires combining such observations with high-resolution soft and hard X-ray spectroscopy---capabilities now available with modern instruments. Instruments such as Reuven Ramaty High Energy Solar Spectroscopic Imager (3~keV--17~MeV; energy resolution $\approx$1~keV, now non-operational; RHESSI \citep{RHESSI}), Solar X-ray Monitor on board Chandrayaan-2 (1.3--15~keV; energy resolution $\approx$175~eV, but lacking high-energy coverage and designed for solar spectroscopy from lunar orbit; XSM \citep{xsm}), and The Spectrometer/Telescope for Imaging X-rays on board Solar Orbiter (4--150~keV; energy resolution $\approx$1~keV, but missing the lower-energy part of the spectrum and operating in a highly elliptical heliocentric orbit; STIX \citep{stix}) do not individually provide continuous broadband X-ray coverage. Therefore, broadband, time-resolved X-ray spectroscopy is essential for simultaneously constraining both the thermal plasma and the accelerated electron distributions. To achieve this, we perform joint spectral fitting X-ray observation of SoLEXS (2--22~keV) and HEL1OS (8--150~keV) onboard \textit{Aditya-L1}, which together provide a continuous energy range of 2--150~keV from the halo orbit around the Sun--Earth Lagrange point~L1. This approach is feasible, as demonstrated previously through the joint fitting of Miniature X-ray Solar Spectrometer (MinXSS) CubeSat and the RHESSI spacecraft \citep{Nagasawa2022} and joint fitting of XSM and STIX spectra \citep{Mithun2024}. 

In this work, we investigate the confined X2.0 flare (SOL2024-10-31T21:20) from NOAA AR~13878 using broad-band X-ray spectroscopy from the Aditya--L1 X-ray spectrometers together with high-resolution EUV and magnetic-field observations from SDO/AIA and SDO/HMI. By closely examining the temporal, spatial, and spectral characteristics of multi-wavelength emissions during the various phases of the confined, long-duration event, the paper investigates the connection between magnetic reconnection processes, the origin of non-thermal and thermal emissions, and the interplay between flare evolution and MFR dynamics. The X-ray spectrometers onboard Aditya-L1 provide continuous broad energy coverage, enabling improved constraints on the low-energy part of the accelerated electron spectrum and, hence, on the partitioning between thermal and non-thermal energies. The AIA and HMI data provide the multi-thermal and magnetic context needed to assess flux-rope activation, loop evolution, and the strength and geometry of overlying fields. A major open question, therefore, is why some long-duration, high-energy flares remain confined despite clear signatures of strong heating, particle acceleration, and flux-rope activation. Addressing, our work presents a comprehensive multi-instrument analysis that: (i) performs joint (i.e. combined data set from SoLEXS and HEL1OS), time-resolved X-ray spectroscopy to quantify plasma heating and particle acceleration; (ii) employs multi-wavelength AIA imaging to track the morphology and dynamics of the suspected flux rope and post-flare arcades; and (iii) examines photospheric magnetic flux evolution to investigate the causes of confinement. The event exhibits two distinct HXR peaks during the impulsive phase, followed by a prolonged, predominantly thermally late phase. The structure of the paper is as follows: Section~\ref{sec:instruments} describes the instruments and data sets; Section~\ref{sec:analysis} presents the multi-wavelength and spectroscopic analysis; Section~\ref{sec:discussion} we discuss observational results; and Section~\ref{sec:conclusions} summarizes the key findings.

%\latex\ \footnote{\url{http://www.latex-project.org/}}

%% The "ht!" tells LaTeX to put the figure "here" first, at the "top" next
%% and to override the normal way of calculating a float position.
%% The asterisk after "figure" tells the compiler to span multiple columns

\section{Instruments and Data} 
\label{sec:instruments}

This study is primarily motivated by the observations of a confined X-class flare from the X-ray spectrometers onboard Aditya-L1 \citep{aditya} —the Solar Low Energy X-ray Spectrometer (SoLEXS; \cite{solex2025}) and the High Energy L1 Orbiting X-ray Spectrometer (HEL1OS; \cite{hel1os2025}). We note that both X-ray spectrometers on board Aditya-L1 are non-imaging spectrometers, optimised for temporal and spectral measurements. We also use data from the Atmospheric Imaging Assembly (AIA; \cite{aia}) and the Helioseismic and Magnetic Imager (HMI; \cite{hmi}) instruments on board the Solar Dynamics Observatory (SDO; \cite{sdo}).

The SoLEXS one of the remote sensing payloads on Aditya-L1, is a sun-as-a-star spectrometer in the energy range of 2 keV to 22 keV with spectral resolution ($\approx$ 170 eV at 5.9 keV) designed to observe soft X-ray emissions from the Sun. The instrument employs a Silicon Drift Detector (SDD) as the sensing element to achieve high-precision energy measurements. To accommodate the extensive dynamic range of solar activity, spanning from A-class to X-class flares, SoLEXS is equipped with two apertures of areas 7.1~mm$^{2}$\ (SSD1) and 0.1~mm$^{2}$
(SSD2).  We note that SDD1 and SDD2 are meant for solar observation in the quiet and active phases of the sun, respectively. In this work, we use SDD2 data for our study. The instrument provides full-spectral data at 1-second cadence, enabling detailed time-resolved studies of rapidly evolving solar X-ray phenomena.

The HEL1OS is the hard X-ray spectrometer onboard the Aditya-L1 Solar Mission, designed to observe high-energy solar emissions associated with flares. It operates over a broad energy range of 8--150~keV, enabling detailed investigations of energetic processes during solar flares. The payload employs two types of detectors to cover its wide operational energy range: a Cadmium Telluride detector operating in the 8--70~keV band and a Cadmium Zinc Telluride detector operating in the 20--150~keV band. The instrument achieves a spectral resolution of approximately 1~keV at 14~keV for CdTe and approximately 7~keV at 60~keV for CZT. The detector crystal dimensions are 5~mm~$\times$~5~mm~$\times$~1~mm for each non-pixelated CdTe detector and 40~mm~$\times$~40~mm~$\times$~5~mm for each CZT detector. The geometric collection area is 0.5~cm$^{2}$ (combined CdTe1 + CdTe2) for the CdTe detectors and 32~cm$^{2}$ (combined CZT1 + CZT2) for the CZT detectors. HEL1OS provides high-temporal-resolution data products, including spectral files for each detector at a cadence of 20~s and light curves for each detector at a cadence of 1~s across multiple energy channels, with an intrinsic time cadence of up to 10~ms, enabling detailed time-resolved studies of dynamic hard X-ray emission during solar flares. Due to ongoing efforts toward calibration and the related data pipeline for the CZT detector, we use measurements only from the CdTe detector of HEL1OS together with SoLEXS data for X-ray spectroscopic analysis. However, to construct X-ray light curves in multiple energy bands, CZT/HEL1OS measurements have been utilized. The detailed procedure of X-ray spectroscopy with Aditya-L1 measurements will be discussed in Section~\ref{sec:joint}.

The AIA records full-disk images of the solar corona and transition region, extending up to $\sim$0.5~$R_{\odot}$ above the photosphere, in multiple EUV and UV passbands. AIA produces narrowband images centred on specific spectral lines corresponding to seven EUV channels: 94~\AA\ (Fe~XVIII), 131~\AA\ (Fe~VIII, Fe~XXI), 171~\AA\ (Fe~IX), 193~\AA\ (Fe~XII, Fe~XXIV), 211~\AA\ (Fe~XIV), 304~\AA\ (He~II), and 335~\AA\ (Fe~XVI). UV observations are obtained at 1600~\AA\ (C~IV) and 1700~\AA\ (near continuum). AIA provides $4096 \times 4096$ pixel images with a spatial scale of $0\farcs6$~pixel$^{-1}$ and a temporal cadence of 12~s for the EUV channels and 24~s for the UV channels. For analysis of AIA images, we employ the SSW routine \texttt{aia\_prep.pro}, which performs the standard geometric and instrumental corrections, aligns the images, and resamples them to a uniform plate scale of $0\farcs6$\,pixel$^{-1}$. In this study, for morphology and kinematical study of the event, we extensively analyse imaging observations from the 94~\AA\ (log~$T$\,=\,6.8), 131~\AA\ (log~$T$\,=\,7.0), 171~\AA\ (log~$T$\,=\,5.8) and 304~\AA\ (log~$T$\,=\,4.7) channels, and to investigate the temporal and spatial characteristic of the flare, the observation in 1600~\AA\ (log~$T$\,=\,5.0) and 1700~\AA\ (log~$T$\,=\,3.7) are also utilize to examine temporal variability of flaring emission in UV.

The HMI provides full-disk measurements of continuum intensity, Doppler velocity, line-of-sight (LOS) magnetic field, and vector magnetic field using the Fe~I 6173~\AA\ absorption line. HMI produces $4096 \times 4096$ pixel images with a spatial scale of $0\farcs5$~pixel$^{-1}$. The temporal cadence is 45~s for intensity, Doppler velocity, and LOS magnetic field, while vector magnetic field data are obtained at a cadence of 720~s. To facilitate a direct comparison between HMI and AIA images, we employ the SSW routine \texttt{hmi\_prep.pro}, which resamples the HMI images from $0\farcs5$~pixel$^{-1}$ to $0\farcs6$~pixel$^{-1}$, matching the plate scale of AIA. In this study, we use intensity and LOS magnetic field data.

In order to discuss the three-dimensional structure of the AR, based on the extrapolation of photospheric magnetograms, a nonlinear force-free field (NLFFF) model is used using the optimisation-based method developed by \cite{Wiegelmann2010, Wiegelmann2012}. As the lower boundary condition, we employed the vector magnetogram obtained from the \texttt{hmi\_sharp\_cea\_720s} data series of HMI/SDO during a pre-flare interval (20:48 UT on 31 October 2024). The magnetogram was rebinned to a spatial sampling of $1.0''$ pixel$^{-1}$ to reduce computational cost. The extrapolation was carried out within a three-dimensional Cartesian domain of $600 \times 256 \times 192$ grid points, corresponding to physical dimensions of approximately $436 \times 186 \times 139$ Mm$^{3}$. In the framework of the NLFFF approximation, the coronal magnetic field is assumed to be force-free, such that the Lorentz force vanishes (i.e., $\mathbf{J} \times \mathbf{B} = 0$), implying that the electric current density $\mathbf{J}$ is parallel to the magnetic field $\mathbf{B}$. In practice, however, numerical implementations of NLFFF extrapolation yield small deviations from this ideal condition. In our case, the angle between $\mathbf{J}$ and $\mathbf{B}$ is found to be $\sim7.7^{\circ}$, which lies well within the acceptable range ($\theta_{J} \lesssim 10^{\circ}$) for reliable NLFFF solutions (e.g., \citealt{DeRosa2015, Mitra2020, Mitra2022}).  
For visualisation and analysis of the three-dimensional magnetic field and derived quantities, we utilised the Visualisation and Analysis Platform for Ocean, Atmosphere, and Solar Researchers (VAPOR; \citealt{Clyne2007}).

\section{Analysis} 
\label{sec:analysis}

We present a detailed multi-instrument analysis of three distinct episodes of hard X-ray (HXR) emission associated with a confined X2.0 solar flare that occurred on 2024 October 31 in Active Region (AR) NOAA 13878. The flare was located at heliographic coordinates N16E25 and originated from a complex magnetic configuration of Hale class $\beta\gamma\delta$. 
After precise dual episodes, HXR emission, the event exhibited a delayed thermal peak. The detailed analysis of these observations is presented in the subsequent subsections.

\subsection{Temporal Phases in X-ray Emissions} 

\begin{figure}[!ht]
    \centering
    \includegraphics[width=1\linewidth]{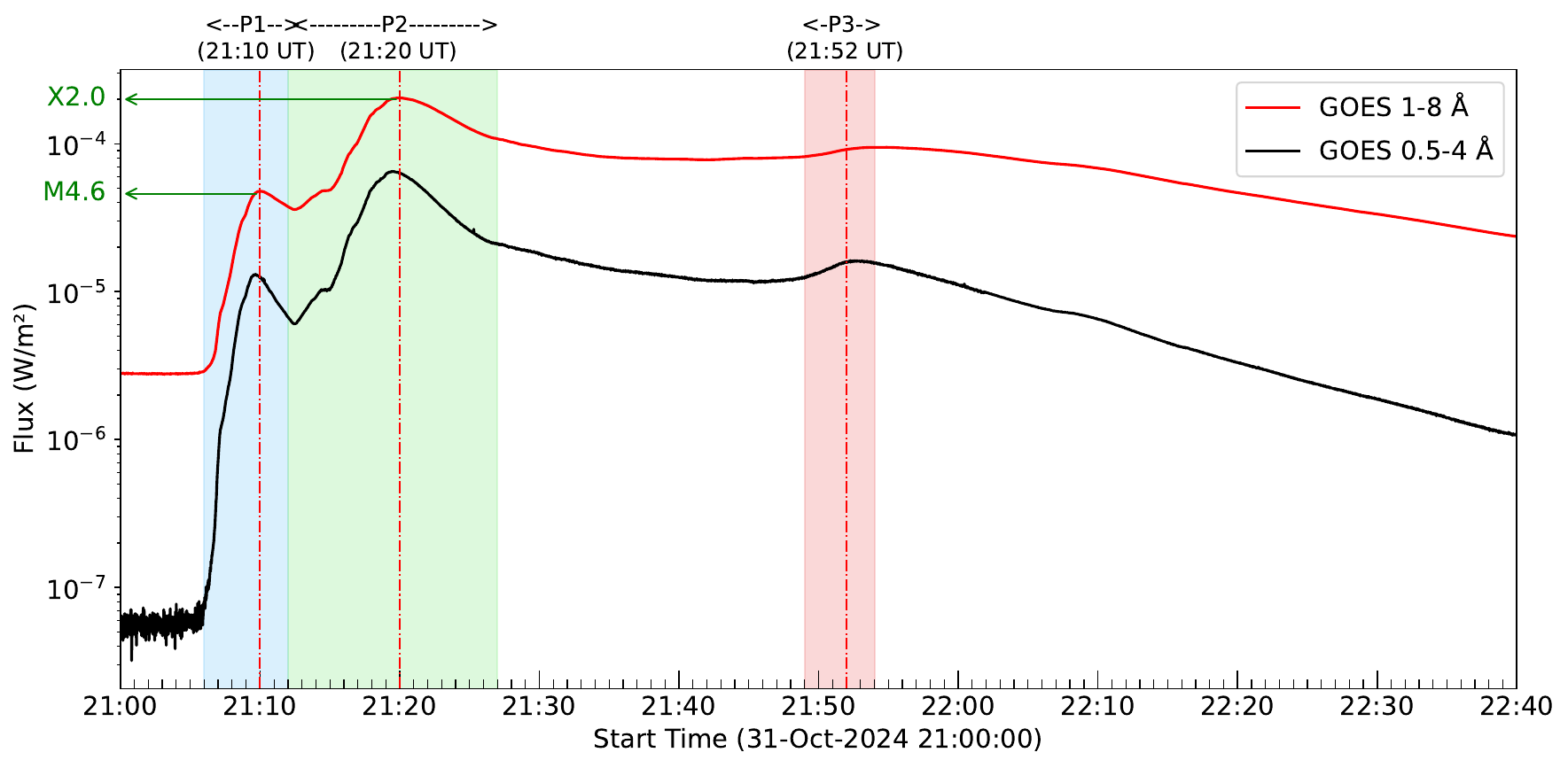}
    \caption{GOES soft X-ray flux profiles on 2024 October 31 from 21:00 UT to 22:40 UT. The red and black curves represent the 1--8~\AA\ and 0.5--4~\AA\ channels, respectively. The blue and green shaded region highlights the main flaring interval corresponding to energy release in AR 13878. The vertical dash-dotted red lines mark the three prominent time instances of peak flux enhancement: P1 (21:10~UT), P2 (21:20~UT), and P3 (21:52~UT).}
    \label{fig:GOES_Xray_Flux_crop1}
\end{figure}

\begin{figure}[!ht]
    \centering
    \includegraphics[width=1\linewidth]{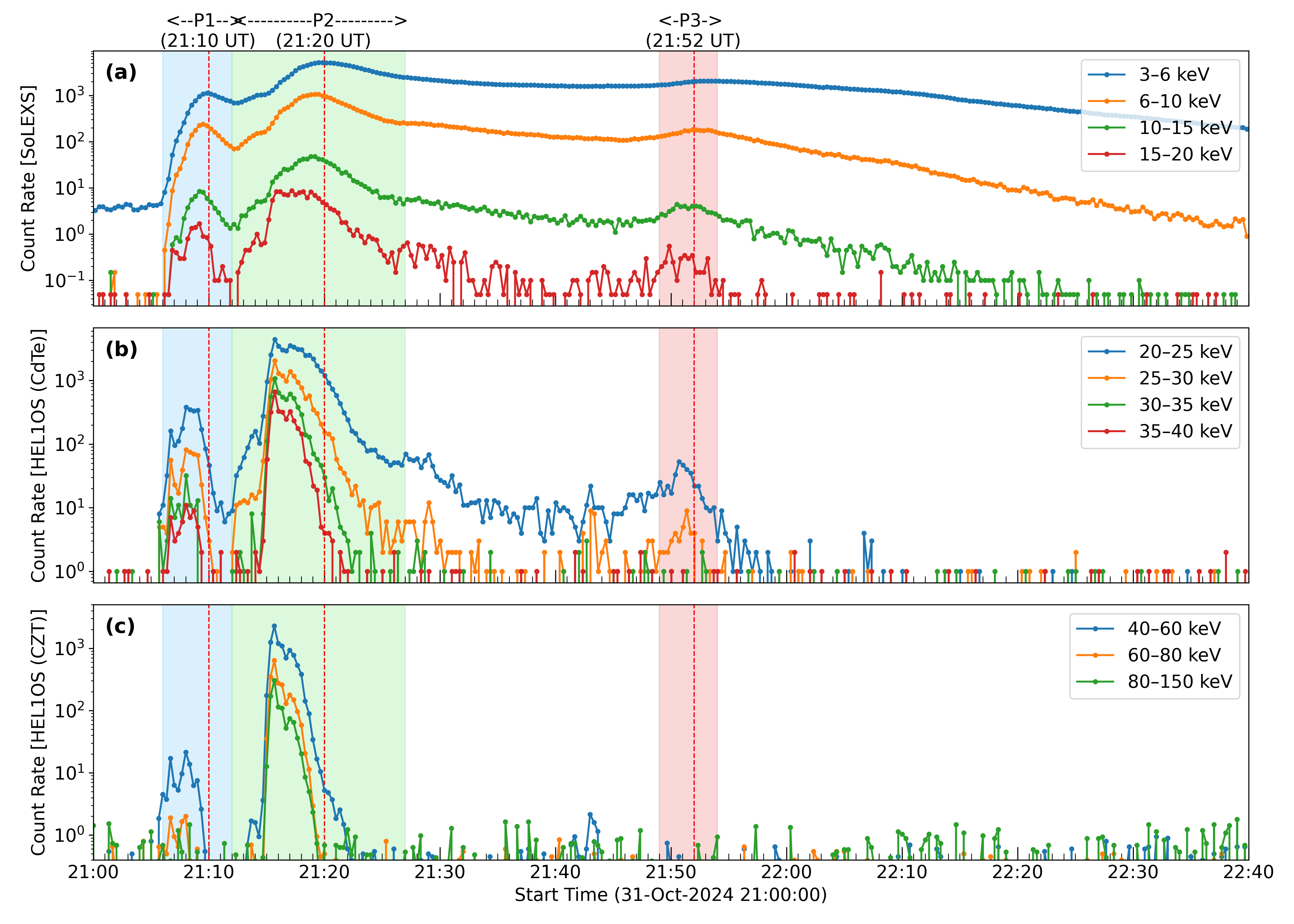}
    \caption{X-ray light curves observed by the X-ray spectrometers onboard Aditya-L1: SoLEXS and HEL1OS. The top panel shows SoLEXS light curves in energy bands of 3--6~keV, 6--10~keV, 10--15~keV, and 15--20~keV. The middle panel shows HEL1OS (CdTe) observations in the 20--25~keV, 25--30~keV, 30--35~keV, and 35--40~keV bands, and the bottom panel shows HEL1OS (CZT) data in the 40--60~keV, 60--80~keV, and 80--150~keV bands. The vertical red dash-dotted lines mark the peak times during P1 (21:10~UT), P2 (21:20~UT), and P3 (21:52~UT).}
    \label{fig:combined1}
\end{figure}

The temporal phases in X-ray emissions observed from the GOES and Aditya-L1 are shown in Figure~\ref{fig:GOES_Xray_Flux_crop1} and Figure~\ref{fig:combined1}, respectively. In Figure~\ref{fig:GOES_Xray_Flux_crop1}, we show the GOES soft X-ray (SXR) time profiles in the 1--8~\AA\ (red) and 0.5--4~\AA\ (black) channels on 2024 October 31 from 21:00 UT to 22:40 UT. Figure~\ref{fig:combined1} presents the X-ray light curves observed by the X-ray spectrometers onboard Aditya-L1 in different energy bands. The top panel shows data from SoLEXS in the energy ranges 3--6~keV, 6--10~keV, 10--15~keV, and 15--20~keV. The middle panel displays observations from HEL1OS (CdTe) in 20--25~keV, 25--30~keV, 30--35~keV, and 35--40~keV energy bands, while the bottom panel shows HEL1OS (CZT) measurements in the 40--60~keV, 60--80~keV, and 80--150~keV energies. The blue, green, and pink shaded regions mark the phase 1 (P1; 21:06-21:12~UT), phase 2 (P2; 21:12-21:27~UT), and phase 3 (P3; 21:49-21:54~UT), respectively. 

We note that P1 and P2 correspond to the primary flaring interval, while P3 marks a mild yet distinct energy release event during the extended decay phase. Clearly, the event exhibits a complex multi-peak temporal structure and characteristics of a long-duration event (LDE). In Figures~\ref{fig:GOES_Xray_Flux_crop1} and \ref{fig:combined1}, the dash-dotted red vertical lines mark epochs of peak flux enhancement during the three phases: P1 (peak at 21:10~UT), P2 (peak at 21:20~UT), and P3 (peak at 21:52~UT). The peak during P2 corresponds to the overall maximum with a magnitude of X2.0, whereas P1 represents a preceding sub-peak of M4.6, occurring approximately 10 minutes before P2. From the X-ray light curves, it is evident that the start and end times of P1 and P2 adequately represent the main energy-release phases of the M- and X-class flares and are consistent with the GOES flare detection criterion based on the 1--8~\AA\ flux variations. Notably, in the present case, we identify an extended decay phase after P2. Such prolonged energy release is not uncommon in LDEs, which frequently exhibit continued emission for several hours beyond the GOES-defined end time (i.e., the time when the 1--8~\AA\ flux decays to halfway between the peak and start flux levels). During this phase, we identify P3 from the subtle yet distinct enhancement in the GOES X-ray flux, with its duration being clearly delineated by the Aditya-L1 X-ray light curves up to 30 keV (see Figure~\ref{fig:combined1}).

The total duration of the event is approximately 1 hour and 40 minutes. The impulsive phase begins near 21:06~UT, and the X-ray flux remains significantly elevated above the pre-flare background level for more than 1 hour and 35 minutes, confirming the long-duration nature of the event. Despite its long-duration behaviour, the flare was not associated with a CME, which is atypical, as most LDEs are typically eruptive in nature.
The dual-peak morphology observed in the GOES SXR profiles suggests two broad episodes of energy release associated with large-scale magnetic reconnection. A delayed third peak over and above the post-flare SXR background implies a secondary process of heating of coronal loops, distinct from the typical post-flare phase. These observations highlight the existence of a non-eruptive, long-duration multi-phase flare, which challenges the conventional understanding of flare–CME association.

The light curves clearly exhibit a dual-peak structure in hard X-rays (HXR). During P1, significant emission is detected up to $\sim$60~keV, whereas P2 exhibits much stronger HXR emission that extends up to $\sim$150~keV. The third enhancement (P3), occurring during the decay phase, is mainly confined to energies $\sim$30~keV, suggesting it to be associated with a weak non-thermal emission. A systematic delay of the peak times with decreasing energy is evident, with lower-energy channels peaking progressively later than the higher-energy channels (see light curve during peak P1 and P2 in Figure \ref{fig:combined1}).  While the GOES data primarily indicate the presence of the third peak in soft X-rays, the \textit{Aditya-L1} observations confirm that P3 is also associated with measurable hard X-ray emission, demonstrating continued high-energy processes during the decay phase of the flare.

\subsection{Temporal Evolution in EUV and UV Channels} 
\label{sec:aia_lc}
\begin{figure}[!ht]
    \centering
    \includegraphics[width=1\linewidth]{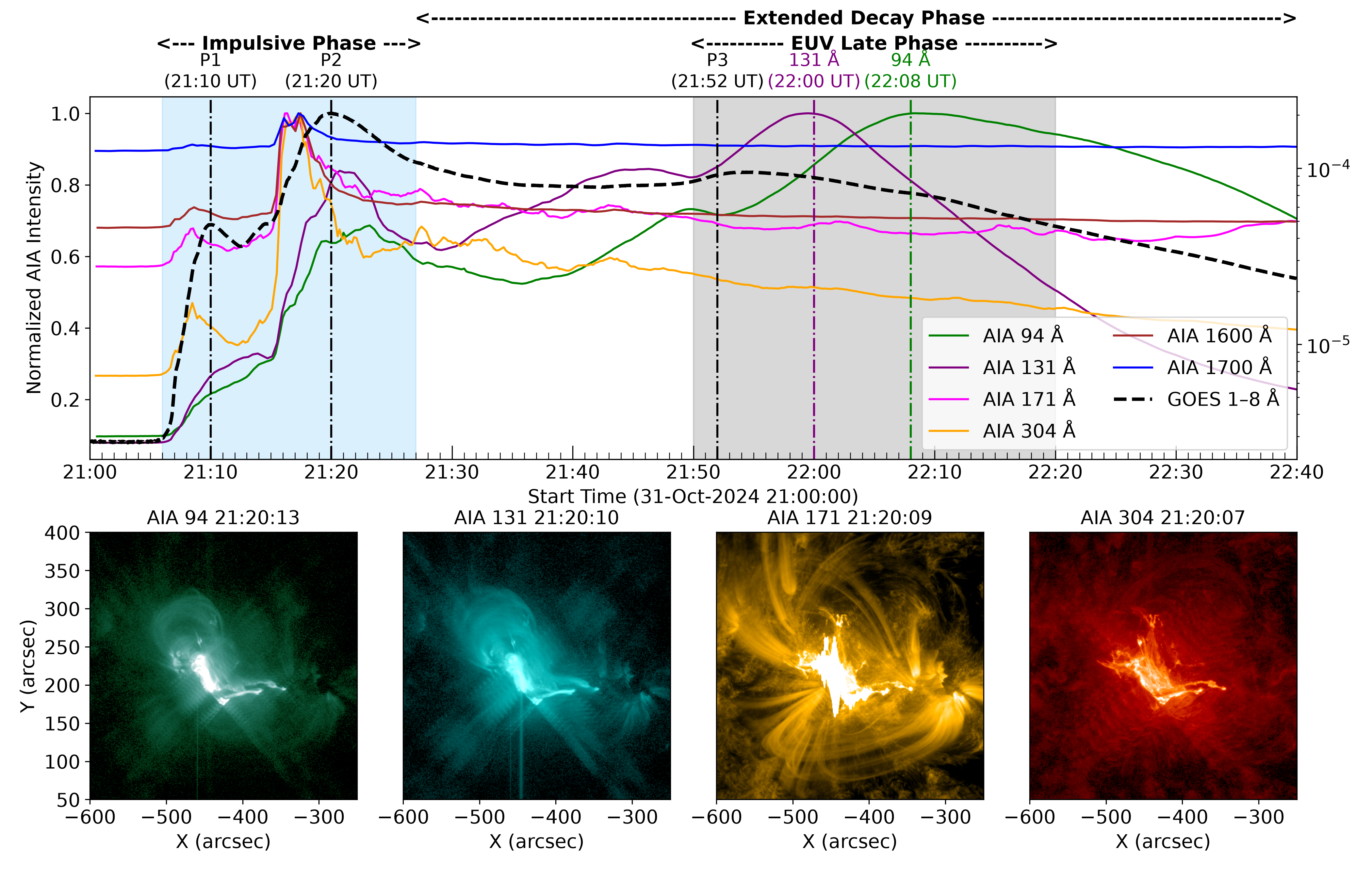}
    \caption{Normalised light curves from SDO/AIA UV and EUV channels (94~\AA, 131~\AA, 171~\AA, 304~\AA, 1600~\AA, and 1700~\AA) together with the GOES 1--8~\AA\ soft X-ray flux (black dashed curve). The blue shaded region (21:06 UT to 21:27 UT) marks the main flaring event, which includes P1 (peak at 21:10~UT) and P2 (peak at 21:20~UT) identified with GOES 1--8 \AA\ flux. This is followed by the extended decay phase. Subsequently, an EUV late phase develops, highlighted by the gray shaded region, and is characterized by a significant enhancement of emission in 131~\AA\ and 94~\AA\ channels of AIA with distinct peaks at 22:00~UT and 22:08~UT, respectively. We also note a subtle enhancement in GOES 1--8~\AA\ flux, peaking at 21:52~UT, marked as P3. The bottom panel shows representative AIA images in different channels at the peak of X2.0 flare in GOES 1--8~\AA\ channel ($\sim$21:20 UT), showing eruptions of a hot channel flux rope structure (94~\AA\ and 131~\AA), the configuration of the overlying loop (171~\AA), and chromospheric ribbons (304~\AA). The animation begins at 21:00 UT and ends at 22:40 UT. The real-time duration of the animation is 24 s. It reveals activation, rapid rise, and subsequent confinement of the hot channel (magnetic flux role) structure.\\     
    (An animation of this figure is available in the online version of the article.)}
    \label{fig:AIA_GOES_combined_lightcurve}
\end{figure}

Figure~\ref{fig:AIA_GOES_combined_lightcurve} shows the normalised multi-wavelength light curves of AR 13878 from SDO/AIA observed in different UV and EUV channels, together with the GOES 1--8~\AA\ soft X-ray flux. The plotted AIA channels include 94~\AA, 131~\AA, 171~\AA, 304~\AA, 1600~\AA, and 1700~\AA. As similar to GOES (Figure \ref{fig:GOES_Xray_Flux_crop1}) and X-ray spectrometers onboard Aditya-L1 (Figure \ref{fig:combined1}) light curves, in Figure \ref{fig:AIA_GOES_combined_lightcurve} also, we mark various temporal structure and instances by blue shaded region (21:06 UT - 21:27 UT) and the vertical dash-dotted lines during P1 (21:10~UT), P2 (21:20~UT), and P3 (21:52~UT). In Figure \ref{fig:AIA_GOES_combined_lightcurve}, we also marked an extended time interval (21:50 UT - 22:20 UT) by a gray shaded region; during this phase, we observe a significant increase in EUV flux in hot coronal channels (i.e. AIA 94~\AA\ and 131~\AA). Notably, the peak intensity of emission (22:00 UT in AIA 131~\AA\ and ~22:08 UT in AIA 94~\AA) reaches even at a significantly higher level than that of the main impulsive peak in these channels. In literature, such kind of enhanced EUV emission in hot and/or warm coronal channels of AIA during the decay phase of the flare has been termed as `EUV late phase' \citep{Woods2011, Dai2018, Li2024a}. In Figure~\ref{fig:AIA_GOES_combined_lightcurve}, we mark the interval of the EUV late phase based on the intensity enhancement observed in the AIA 131~\AA\ and 94~\AA\ light curves. This identification is further confirmed through a close inspection of the corresponding AIA 131~\AA\ and 94~\AA\ images (see also Figure~\ref{fig:box_aia}; Section~\ref{sec:aia_phase2}). On the other hand, cooler EUV and UV channels (AIA 171~\AA, AIA 304~\AA, 1600~\AA, and 1700~\AA) do not show enhancement.

\subsection{Coronal dynamics during impulsive energy release (21:06 UT--21:27 UT)}
\label{sec:ph1}
\begin{figure}[!t]
    \centering
    \includegraphics[width=0.8\linewidth]{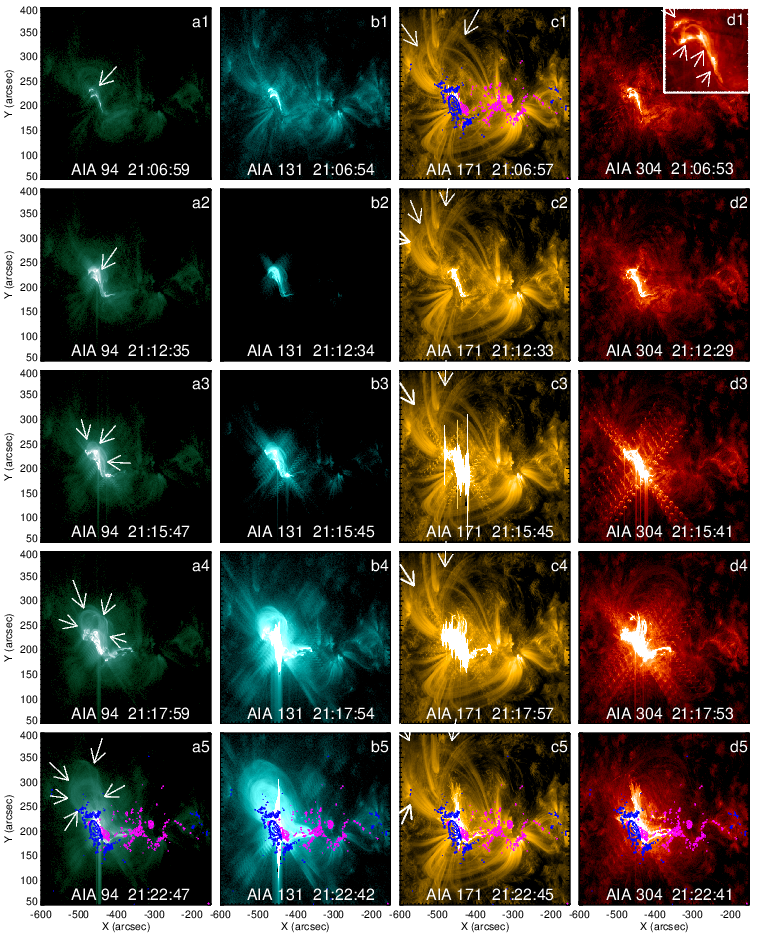}
    \caption{Multi-wavelength SDO/AIA images showing the temporal evolution of the flare during the impulsive phase. The panels display observations in the 94~\AA, 131~\AA, 171~\AA, and 304~\AA\ channels. Brightening of the core region and the expansion of the magnetic flux rope are indicated by white arrows in panels a1--a5, while the expansion of the overlying coronal loops is marked by white arrows in panels c1--c5 during both the P1 and P2 phases, indicating rapid magnetic reconnection and intense plasma heating. In panels (c1) and (a5–d5), we overlay the magnetic-field contours-- blue for negative polarity and magenta for positive polarity. The inset in panel (d1) presents the zoomed view of the core region, showing the preflare localised brightenings or kernels (by arrows).}
    \label{fig:aia_phase1}
\end{figure}

Figure~\ref{fig:aia_phase1} illustrates the morphological evolution of the flare during the impulsive phase using multi-wavelength SDO/AIA observations in the 94~\AA, 131~\AA, 171~\AA, and 304~\AA\ channels at selected times between 21:06~UT and 21:23~UT. These images capture the dynamic development of the flare during the first two energy release phases (P1 and P2), corresponding to the dual-peak structure observed in the X-ray light curves (see Figure \ref{fig:GOES_Xray_Flux_crop1} and \ref{fig:combined1}). The pre-flare phase is characterized by compact loop brightenings observed in the coronal 94~\AA, 131~\AA, and 171~\AA\ channels (a1–c1), together with bright flare kernels in the chromospheric 304~\AA\ channel (d1 and inset), all located near the polarity inversion line (PIL). During the early impulsive phase ($\approx$21:08~UT onward until the peak at 21:10 UT), localized brightenings are observed in the hot coronal channels of 94~\AA\ and 131~\AA\ as seen in panels a2--a3 and b2--b3. At this time, the MFR structure is compact and embedded within the core region (marked by an arrow in panels a1 and a2). The AIA 94~\AA\ images reveal a well-defined MFR, while the overlying coronal loops are clearly visible in the 171~\AA\ channel toward the north-east direction.
A gradual expansion of the MFR is observed from $\approx$ 21:15 UT onward (marked by multiple white arrows in panels a3--a5). We note that MFR appears to emanate from the core region ($\approx$ 5 minutes) after peak P1. We also observe a gradual expansion in the overlying coronal loops in AIA 171~\AA\ channel. The expansion of these coronal loops is visible in the image sequence (marked by white arrows in panels c1--c5). These overlying loops act as a restraining magnetic arcade, constraining the eruption of the flux rope. The observation from AIA 131 \AA\ and 94 \AA\ provides a consistent picture about the MFR expansion from the core. Unlike the other coronal channels (AIA 94~\AA, 131~\AA, and 171~\AA), the AIA 304~\AA\ images show intense brightening close to the core region. In panels (c1) and (a5–d5), we overlay the magnetic-field contours-- blue for negative polarity and magenta for positive polarity. These observations indicate that the chromospheric response to the coronal energy release is largely localised and confined near the PIL.

\begin{figure}[!ht]
    \centering
    \includegraphics[width=0.8\linewidth]{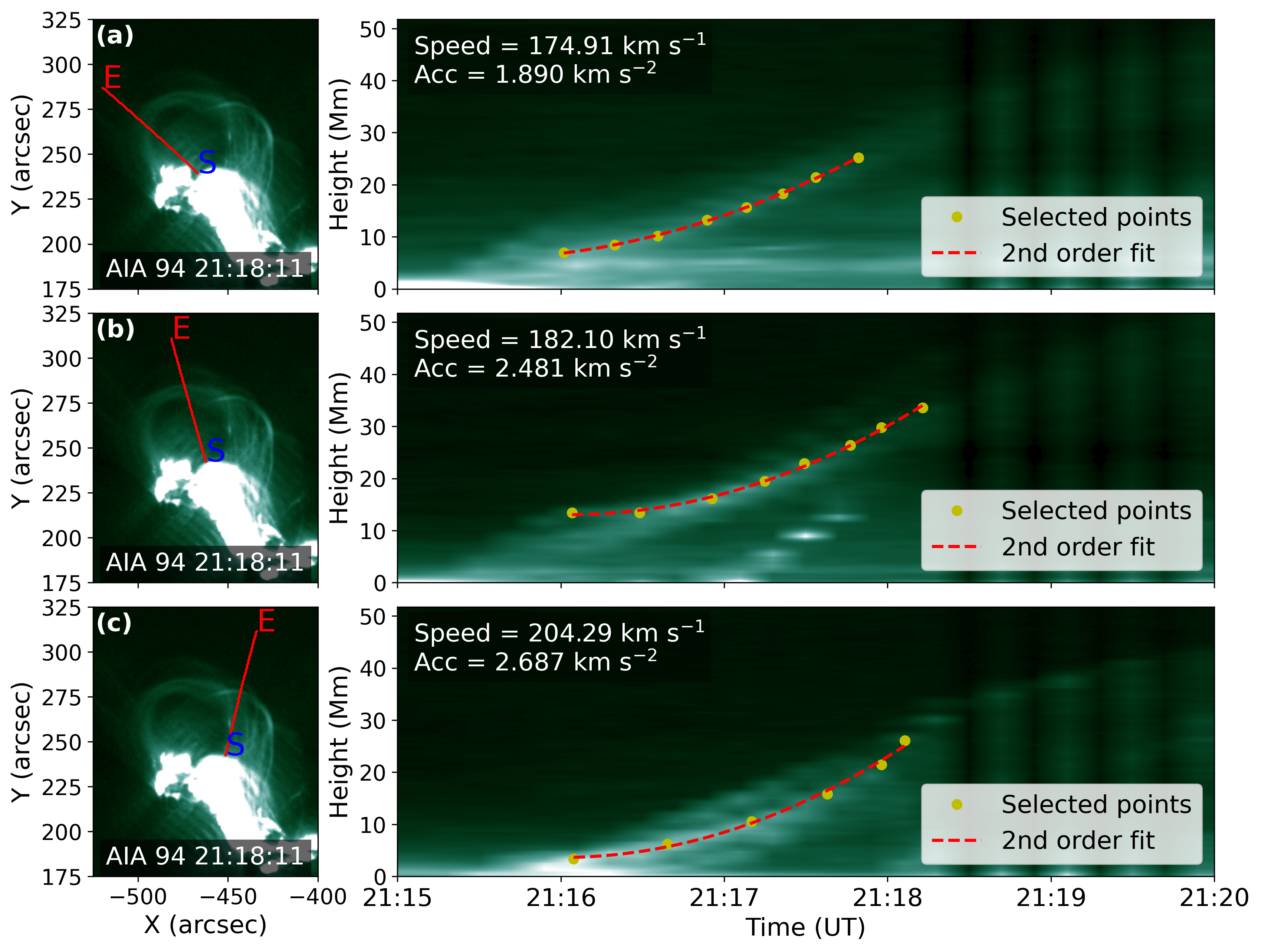}
    \caption{Height--time analysis of the magnetic flux rope (MFR) observed in the AIA 94~\AA\ channel. Left panels (a--c) show the locations of the three slits used for time--distance analysis. The right panels display the corresponding height-time maps, with yellow dots indicating selected feature-tracking points and red dashed lines showing second-order polynomial fits. The analysis reveals that velocities and accelerations indicate an accelerating rise of the MFR during the impulsive phase.}
    \label{fig:94_slice_pair}
\end{figure}

Figure~\ref{fig:94_slice_pair} shows the height--time evolution of the magnetic flux rope (MFR) observed in the AIA 94~\AA\ channel along three different slit positions during the impulsive phase of the flare. The slit locations are indicated in the left panels (a--c) by straight line S-E, while the right panels show the corresponding time-slice diagram. From the time-slice diagram, we clearly notice the expansion of the leading edge of the hot channels along the corresponding direction, specified by the slit. The yellow symbols show a few manually selected points along the trajectory of the flux rope, while the red dashed curve shows a second-order polynomial fit to height-time data. We find upward expansion of MFR , with derived plane-of-sky velocities of approximately 174.9~km~s$^{-1}$, 182.1~km~s$^{-1}$, and 204.2~km~s$^{-1}$, and accelerations of about 1.89~km~s$^{-2}$, 2.48~km~s$^{-2}$, and 2.68~km~s$^{-2}$, respectively, along the 3 slit. These results indicate a non-uniform and accelerating rise of the MFR during the impulsive phase.

\subsection{X-ray Spectroscopy}
\subsubsection{Joint spectral fitting of SoLEXS and HEL1OS/CdTe data}
\label{sec:joint}

X-ray spectra of the solar flare observed with SoLEXS and HEL1OS are analysed to derive the thermal and non-thermal plasma parameters. Although OSPEX\footnote{\url{https://hesperia.gsfc.nasa.gov/ssw/packages/spex/doc/ospex_explanation.htm}} \citep{ospex}, a part of the SolarSoft package, is widely used for modelling X-ray spectra when data from a single instrument is analysed, e.g. RHESSI, STIX, it currently lacks the capability to perform simultaneous multi-instrument spectral fitting with a common model. Therefore, to conduct joint spectral fitting of SoLEXS and HEL1OS, we employ the XSPEC\footnote{https://heasarc.gsfc.nasa.gov/docs/software/xspec/} spectral analysis package \citep{xspec}, which supports combined fitting of spectra from multiple instruments \citep{Nagasawa2022, Mithun2024}. In particular, we make use of its Python interface, PyXSPEC. Time-resolved X-ray spectra are generated for both SoLEXS and HEL1OS covering the duration of the flare. While SoLEXS provides data at a uniform cadence of 1~s, HEL1OS records spectra nominally at 20~s cadence. To achieve sufficient count statistics in SoLEXS data, longer integration intervals are required. Thus, we adopt a uniform cadence of 20~s for the generation of spectra for both SoLEXS and HEL1OS for the same time intervals for consistency.

SoLEXS spectra for the selected flare intervals are produced using the
\texttt{solexs-genmultispec} task from the SoLEXS Tools Software v1.1,
along with the latest calibration database. The current calibration does not incorporate the dead-time correction; however, this affects only the overall normalisation of the spectra, not the other plasma parameters. The non-solar background for SoLEXS is derived from early mission observations performed with the aperture door closed, as well as from periods during which the spacecraft observed a quiet X-ray sky. This background spectrum is incorporated during the XSPEC fitting. HEL1OS spectra are generated for the same time intervals using the latest calibration database (CALDB version \texttt{CAL\_epoch20231001}). For HEL1OS, the pre-flare intervals are used to construct the background spectrum. The background-subtracted spectra from both instruments are then used for the joint spectral fitting.

Modelling the flare X-ray spectra in XSPEC requires components that describe the thermal and non-thermal emission. For the thermal contribution, we use the
chisoth\footnote{https://github.com/xastprl/chspec} local model, which is based on the CHIANTI atomic database and
accounts for continuum and line emission from an optically thin, ionised plasma \citep{Mondal2021}. The model parameters include temperature, emission measure, and elemental abundances. The non-thermal emission arises from bremsstrahlung produced by accelerated electrons, follows a power-law energy distribution \citep{kepa2008, Kontar2011}. The resultant photon spectrum can be approximated with a broken power law model with a flat spectral index below the break energy. Thus, we use a model chisoth+broken power law to cover the thermal and non-thermal components.

Although SoLEXS provides spectroscopic coverage in the 2--22~keV energy range, the instrumental response between 2.0 and 2.8~keV is currently not well modelled. Therefore, the spectrum above 2.8~keV is fitted while the upper energy limit is determined based on the signal-to-noise ratio for the spectrum of each time bin. As discussed in the section~\ref{sec:instruments}, the HEL1OS employs two types of detectors, CdTe (8-70 keV) and CZT (20-150 keV). However, at present, the HEL1OS/CZT instrument response has not been finalised, and efforts are underway to optimise and verify it. Considering this, for X-ray spectroscopy of this event, we use SoLEXS and HEL1OS/CdTe data only. HEL1OS/CdTe provides spectroscopic coverage in the 8--70~keV energy range, the instrumental response between 8.0 and 9.0~keV is currently not well modelled. Therefore, the spectrum above 9.0~keV is fitted with the upper energy limit again is determined based on the signal-to-noise ratio for the spectrum of each time bin. Further, to account for calibration uncertainties in the SoLEXS and HEL1OS/CdTe, we incorporate a model systematic uncertainty of 5\%, which provides statistically acceptable fits.

As we present the joint spectral fitting results from the SoLEXS and HEL1OS/CdTe instruments, it is useful to briefly describe the methodology. In astrophysical X-ray spectroscopy, the observed count spectra cannot be directly inverted to obtain the photon flux spectrum; therefore, a forward-folding approach is employed (see the XSPEC manual for details; \url{https://heasarc.gsfc.nasa.gov/docs/software/xspec/manual/node10.html}). A physical model (here, an isothermal plasma emission model combined with a broken power-law model for non-thermal emission) is defined by a set of parameters, convolved with the instrument response, and fitted to the observed count spectra from both instruments. The instrument response incorporates various aspects of the instrument calibration. The model parameters are varied to obtain the best fit to the observed data.

\begin{figure}
    \centering
    \includegraphics[width=1\linewidth]{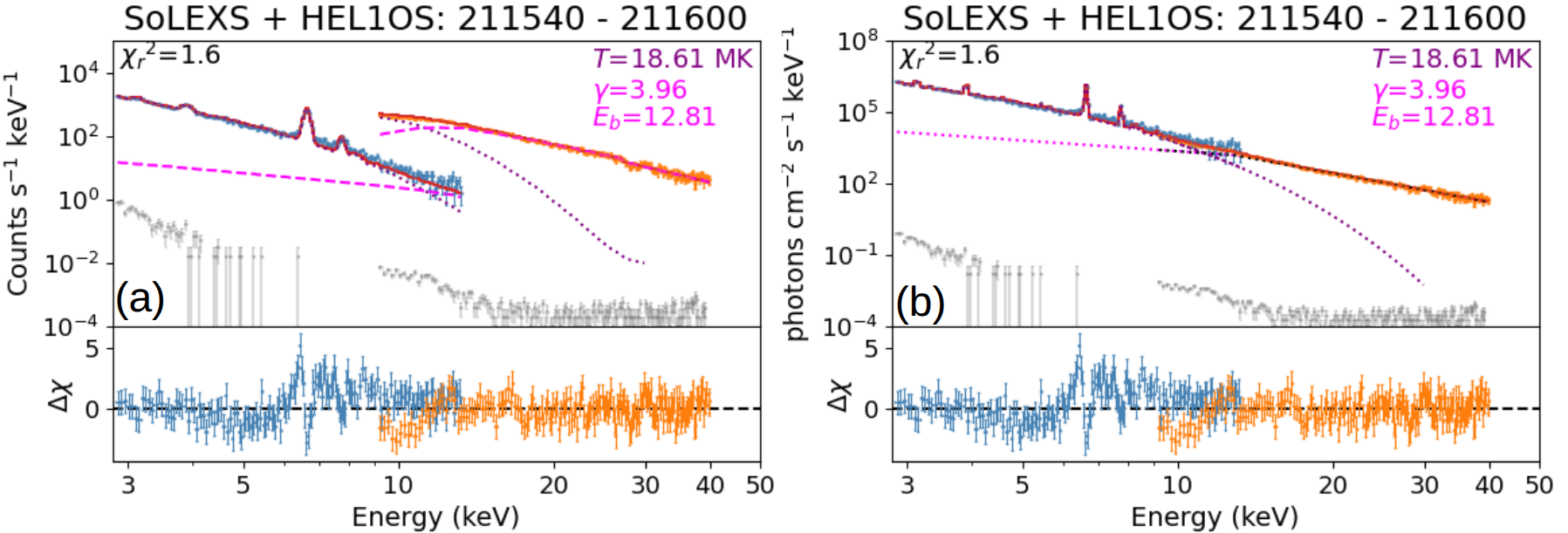}
    \caption{Representative joint spectral fitting of SoLEXS and HEL1OS/CdTe observations during the interval 21:15:40--21:16:00 UT. Panel (a) shows the folded count spectrum, where the observed count rates from SoLEXS (blue) and HEL1OS/CdTe (orange) are fitted simultaneously with the detector response-convolved model in units of counts s$^{-1}$ keV$^{-1}$. Panel (b) presents the corresponding unfolded photon spectrum in physical units, where the y-axis represents the photon flux in units of photons cm$^{-2}$ s$^{-1}$ keV$^{-1}$. The purple dotted and magenta dashed curves denote the thermal and non-thermal model components, respectively, while the red curve represents their sum. The lower sub-panels show the fit residuals ($\Delta\chi$). It can be seen that in physical photon flux units, the data from two instruments match well and show no discontinuity.}
    \label{fig:uf}
\end{figure}

Figure~\ref{fig:uf}(a) shows the observed count spectra (counts~s$^{-1}$~keV$^{-1}$) from the two instruments, overplotted with the corresponding convolved models that provide the best joint fit during the HXR peak phase. Because the effective areas of the two instruments differ, the observed count spectra are expected to differ even for the same physical model, resulting in the apparent discontinuity in count rates. However, this difference is fully accounted for through the respective instrument response matrices, allowing the same physical model to be consistently fitted to both datasets.

To illustrate this, Figure~\ref{fig:uf}(b) shows the corresponding unfolded spectra with the models overplotted on the approximately inverted photon spectra. In physical photon flux units (photons~cm$^{-2}$~s$^{-1}$~keV$^{-1}$), the data from the two instruments agree well and show no discontinuity. Since the unfolding process is only approximate, we present the folded count spectra with the convolved models in the subsequent detailed joint spectral fitting results, as this provides a more rigorous and appropriate representation of the analysis.

\begin{figure}[!ht]
    \centering
    \includegraphics[width=1\linewidth]{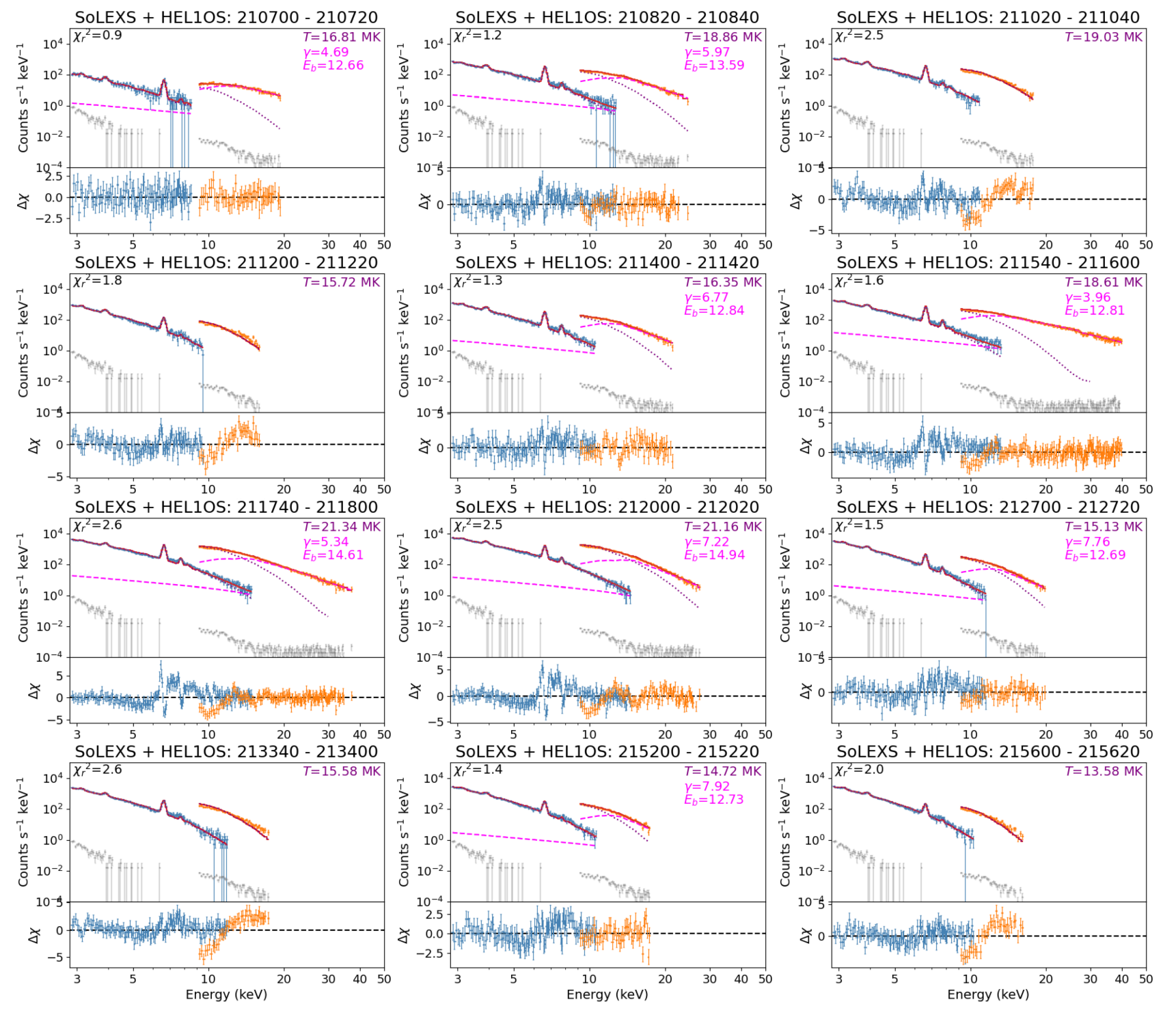}
    \caption{Joint spectral fitting of time-resolved X-ray spectra using SoLEXS and HEL1OS/CdTe observations during two distinct episodes of hard X-ray (HXR) emission in the confined X2.0 solar flare and during its decay phase. The observed spectra from SoLEXS (blue) and HEL1OS/CdTe (orange) are shown, along with their corresponding residuals in the lower panels using matching colours. The best-fit model consists of a chisoth thermal component (purple dotted line) and a broken power-law non-thermal component (magenta dashed line), while the combined (chisoth $+$ broken power-law) model is shown as a red stepped line. The best-fit spectral parameters and the reduced chi-square ($\chi^{2}_{\mathrm{r}}$) values are indicated in each panel for representative time intervals, demonstrating the quality of the combined thermal and non-thermal spectral fits.}   
    \label{fig:aditya_joint}
\end{figure}

\subsubsection{Evolution of thermal and non-thermal characteristics}

In Figure~\ref{fig:aditya_joint}, we present a few representative time-resolved X-ray spectra showing combined measurements from SoLEXS and HEL1OS/CdTe along with joint spectral fitting. We have performed the joint spectral fitting with a 20-second integration time over consecutive intervals from 21:07 UT to 22:20 UT, spanning over two distinct episodes of hard X-ray (HXR) emission and the decay phase of the confined X2.0 solar flare. The observed spectra from SoLEXS (blue) and HEL1OS/CdTe (orange) are shown together with their corresponding residuals in the lower panels in matching colours. The best-fit spectral model consists of a thermal component represented by the chisoth model (purple dotted line) and a non-thermal component modelled using a broken power-law (magenta dashed line), while the combined (\texttt{chisoth} $+$ broken power-law) model is displayed as a red stepped line. The best-fit spectral parameters and the reduced chi-square ($\chi^{2}_{\mathrm{r}}$) values are indicated in each panel for a representative time interval, illustrating the quality and consistency of the joint thermal and non-thermal spectral fits.

\begin{figure}[!ht]
    \centering
    \includegraphics[width=0.8\linewidth]{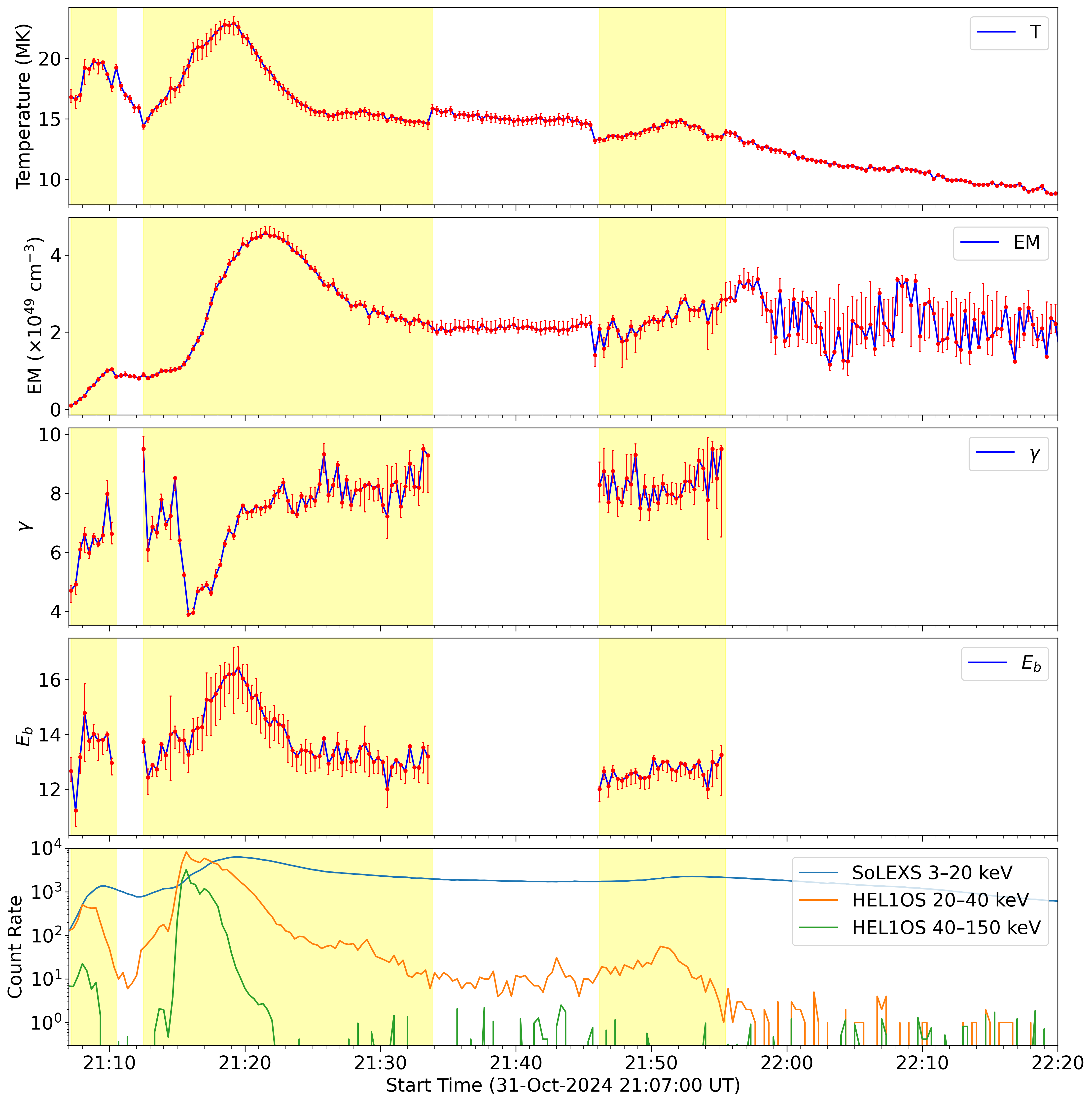}
    \caption{Time evolution of spectral fitting parameters derived from X-ray observations. From top to bottom: plasma temperature ($T$), emission measure (EM), photon spectral index ($\gamma$), and break energy ($E_{b}$). The bottom panel shows observed count rates from SoLEXS (3--20~keV, blue), HEL1OS 20--40~keV (orange), and HEL1OS 40--150~keV (green). The yellow shaded regions mark intervals dominated by non-thermal emission. 
    Error bars correspond to one-sigma uncertainties calculated by MCMC analysis.}
    \label{fig:combined_parameter}
\end{figure}

Figure~\ref{fig:combined_parameter} shows the temporal evolution of the spectral parameters derived from time-resolved X-ray fitting during the flare. From top to bottom, the panels display the plasma temperature ($T$), emission measure (EM), photon spectral index ($\gamma$), and break energy ($E_{b}$), together with the observed count rates in different energy bands. One-sigma errors from Markov Chain Monte Carlo (MCMC) analysis are also shown for each spectral parameter.

During the two distinct episodes of hard X-ray (HXR) emission in the impulsive phase of a confined X2.0 solar flare, we observe three phases of non-thermal emission, shaded in the yellow region. The plasma temperature peaks for the first phase at $\sim$19 MK, the second phase peak at $\sim$22 MK, and the third decay phase at $\sim$15 MK, respectively, coincide with the rise of hard X-ray flux. Simultaneously, the emission measure exhibits a gradual increase during the first phase and a sharp increase during the second phase. After 21:52 UT, we notice a further increase in peaks. It is interesting to note that after this peak, temperature decreases and emission measures increase. 
Since the spectral fitting employs a composite model consisting of \texttt{chisoth} and a broken power-law component, the broken power-law is included only during time intervals that show clear non-thermal emission. During intervals dominated by thermal emission, the broken power-law component is frozen. This treatment results in an apparent sudden increase in temperature and a corresponding decrease in emission measure immediately after the yellow-shaded interval for both non-thermal phases. The photon spectral index decreases (indicating spectral hardening) during periods of non-thermal emission and reaches minimum values during these intervals, consistent with efficient acceleration of high-energy electrons. The first episode is characterised by relatively weak non-thermal emission compared to the second episode, as indicated by the photon index varying from $\sim$4.34 to $\sim$7.94 during the first phase, whereas during the second phase it varies more strongly from $\sim$3.96 to $\sim$9.5. The predominance of thermal emission is observed during the third (decay) phase, with the photon index varying between $\sim$7.27 to $\sim$9.5.

We find that the break energy, $E_{b}$, varies between $\sim$11.5~keV to $\sim$14.5~keV during the first non-thermal phase and increases from $\sim$12~keV to $\sim$16~keV during the impulsive phase and from $\sim$13~keV to $\sim$15~keV during the decay phase. The count-rate profiles from SoLEXS (3--20~keV) and HEL1OS (20--40~keV and 40--150~keV) closely track the evolution of the thermal and non-thermal components. The persistence of elevated thermal parameters after the decay of hard X-ray flux supports the presence of prolonged heating during the late phase of the event.

\subsection{Origin of enhanced heating during extended decay phase (21:27 UT--22:40 UT)}
\label{sec:aia_phase2}

\begin{figure}[!ht]
    \centering
    \includegraphics[width=0.8\linewidth]{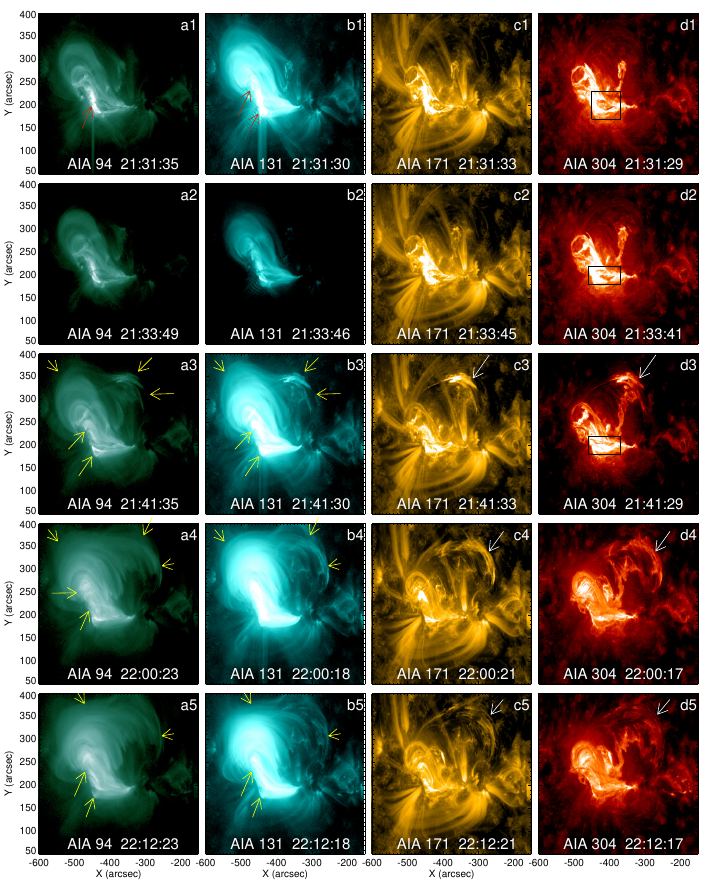}
    \caption{Multi-wavelength SDO/AIA images showing the late-phase evolution of the flare between 21:31~UT and 22:12~UT in the 94~\AA, 131~\AA, 171~\AA, and 304~\AA\ channels. The panels reveal continued brightening and expansion of the core region, marked by the red arrow in panels a1 and b1, as well as the formation and lateral expansion of a large coronal arcade, indicated by the yellow arrows in panels a3, a4, a5, and b3, b4, and b5. Downflows of cooling plasma are visible in the 171~\AA\ and 304~\AA\ channels, marked by white arrows in panels c3--c5 and d3--d5. Activation of nearby filamentary structures is observed in panels d1, d2 and d3, highlighted by a black box.}
    \label{fig:aia_phase2}
\end{figure}

Figure~\ref{fig:aia_phase2} presents the morphological evolution of the flare during the extended decay phase using multi-wavelength SDO/AIA observations in the 94~\AA, 131~\AA, 171~\AA, and 304~\AA\ channels. After the impulsive phase, the flaring region undergoes continued brightening and gradual expansion, most prominently observed in the hot coronal channels of 94~\AA\ and 131~\AA. In Figure~\ref{fig:aia_phase2}, expansion of the core region is marked by the red arrow in panels a1 and b1, while post-eruptive loops and arcades are highlighted by yellow arrows in panels a3--a5 and b3--b5.
Another interesting characteristic of the decay phase is the downward-moving plasma structures, which are observed along the post-flare loops in the 171~\AA\  and 304~\AA\ images. These downflows are marked by white arrows in panels c3--c5 and d3--d5, indicating plasma drainage channelised by magnetic loops. In the AIA 304~\AA\ channel, an interesting feature appears in the form of an evolving cool filament structure at the core region; this structure is shown in panels (d1--d3 in the black box).

\begin{figure}[!ht]
    \centering
    \includegraphics[width=0.8\linewidth]{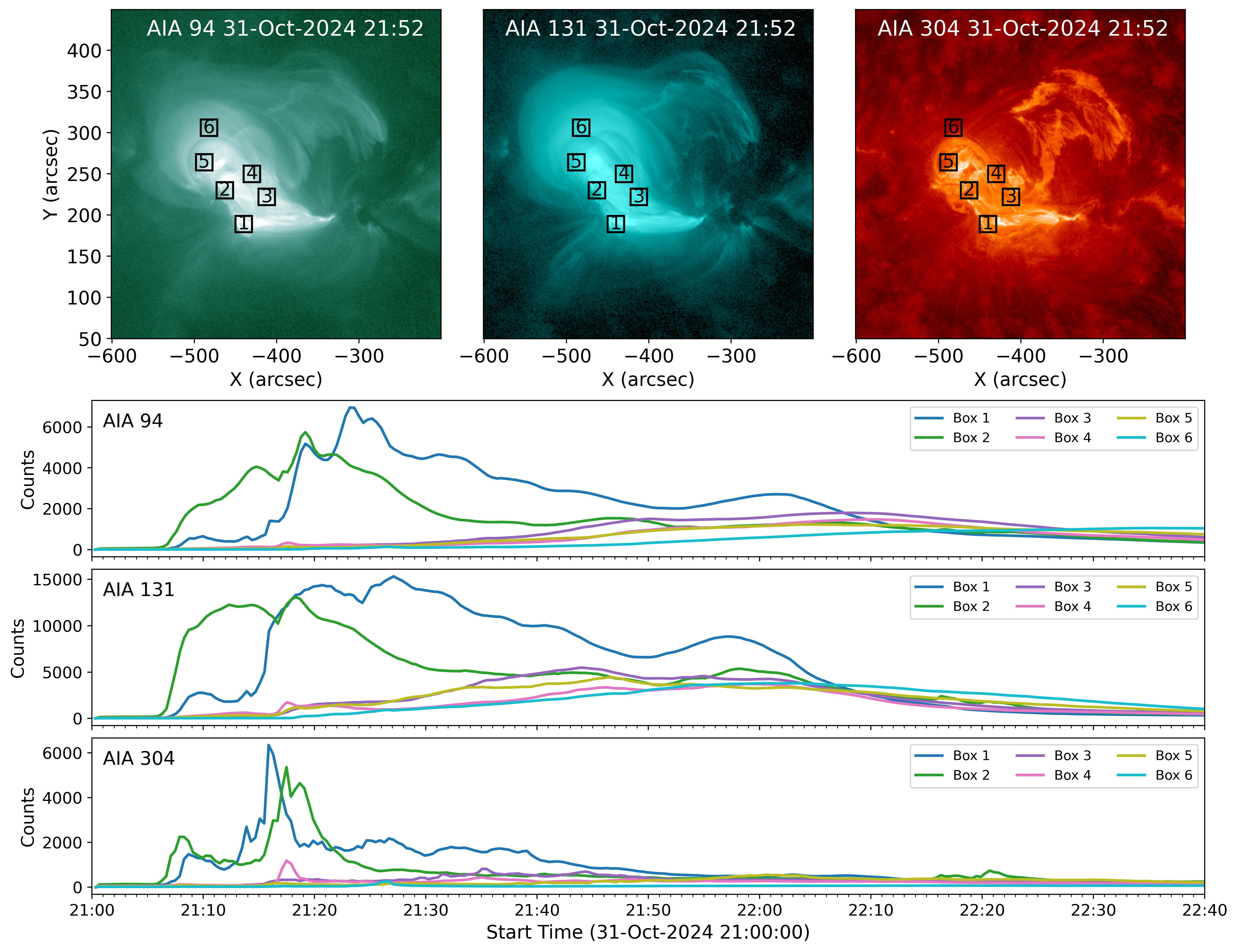}
    \caption{Top panels: SDO/AIA images in the 94~\AA, 131~\AA, and 304~\AA\ channels at P3 (21:52~UT), showing the locations of selected analysis boxes (each of size $20 \times 20$~arcsec$^{2}$). Bottom panels: temporal evolution of the integrated intensities for six representative regions. Boxes~1 and~2 sample the compact core region and are dominated by emission during Phase~1 and Phase~2, while boxes~3--6 trace the flare loops and primarily contribute during EUV late phase.}
    \label{fig:box_aia}
\end{figure}

To investigate the spatial origin of the third-phase X-ray peak (P3 at 21:52~UT; see in Figures \ref{fig:GOES_Xray_Flux_crop1} and \ref{fig:combined1}) and EUV late phase emission (see Figure \ref{fig:AIA_GOES_combined_lightcurve}) observed during 21:50~UT to 22:20~UT, a set of 12 regions (each of size $20 \times 20$~arcsec$^{2}$) were defined across the flaring area. The corresponding light curves were constructed to identify which locations contribute most significantly to the observed integrated flux variability during this phase. Figure \ref{fig:box_aia} shows the spatially resolved intensity evolution of the flare observed in the AIA 94~\AA, 131~\AA, and 304~\AA\ channels. We show only six representative boxes out of 12 boxes, as they are sufficient to characterise the spatial distribution and temporal evolution of the emission. The top row shows the AIA images with the selected six boxes, while the bottom row displays the light curves corresponding to these regions. The light curve analysis reveals that boxes~1 and~2 correspond to the core flare region and capture emission associated with the impulsive phase, gradual phase, along with EUV late phase. On the other hand, boxes~3--6 sample the coronal loop structures that evolve after the impulsive phase, which includes X-ray emission during P3 and an extended late EUV phase. The light curve further reveals that the EUV late phase exhibits the strongest enhancement in the hot coronal channels AIA 94~\AA\ and AIA 131~\AA\ from outer overlying coronal loop systems, while it is almost fully absent in the cool channel of AIA 304~\AA.

\begin{figure}[!ht]
    \centering
    \includegraphics[width=0.8\linewidth]{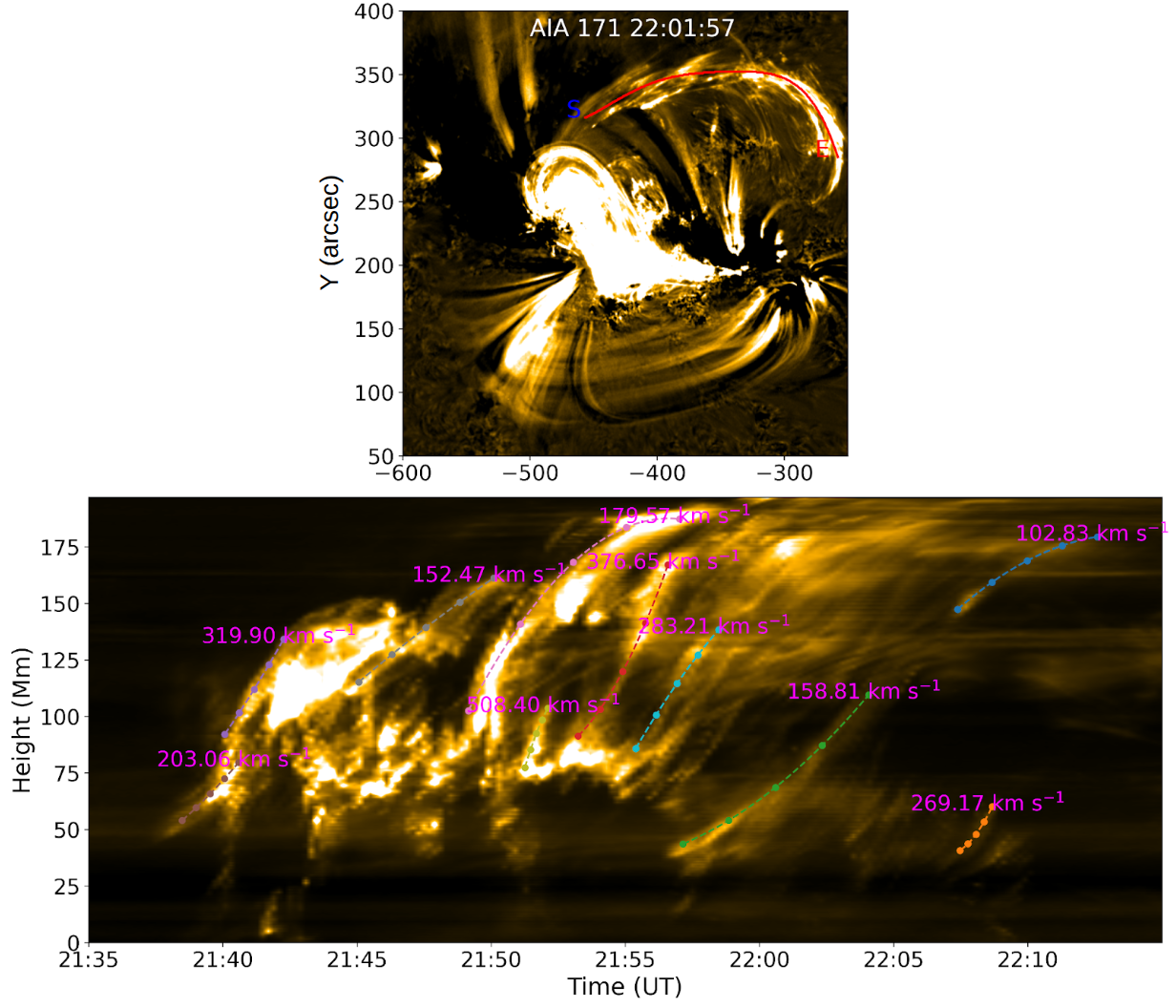}
    \caption{Top panel: SDO/AIA 171~\AA\ image at 22:01:57~UT showing the curved slit (red line) used for height--time analysis, with the start (S) and end (E) points marked. Bottom panel: corresponding height--time map along the slit. Colored dotted lines indicate tracked intensity patterns, and the associated plane-of-sky velocities are labelled in km~s$^{-1}$.}
    \label{fig:171_flow}
\end{figure}

An interesting feature observed in AIA 171 \AA\ images is the flow of plasma along the overlying field lines. In Figure~\ref{fig:171_flow} presents a quantitative analysis of coronal plasma downflow observed in the AIA 171~\AA\ channel (Figure~\ref{fig:aia_phase2} c3--c5 and d3--d5). The top panel shows an AIA 171~\AA\ image at 22:01:57~UT, in which a curved slit (red line) is placed along the trajectory of the loop-like coronal structure, with the start (S) and end (E) points marked. The bottom panel shows that the corresponding time-slice constructed along the slit which captures many instances of moving plasma downflow. Multiple flow patterns are identified and tracked, and their trajectories are fitted with a second-order polynomial function (see dashed colored lines). The derived plane-of-sky velocities range from approximately 100~km~s$^{-1}$ to more than 500~km~s$^{-1}$, as annotate in Figure~\ref{fig:171_flow}.

\subsection{Magnetic Structure of the Active Region}
\subsubsection{Evolution of photospheric magnetic field}

\begin{figure}[!ht]
    \centering
    \includegraphics[width=0.8\linewidth]{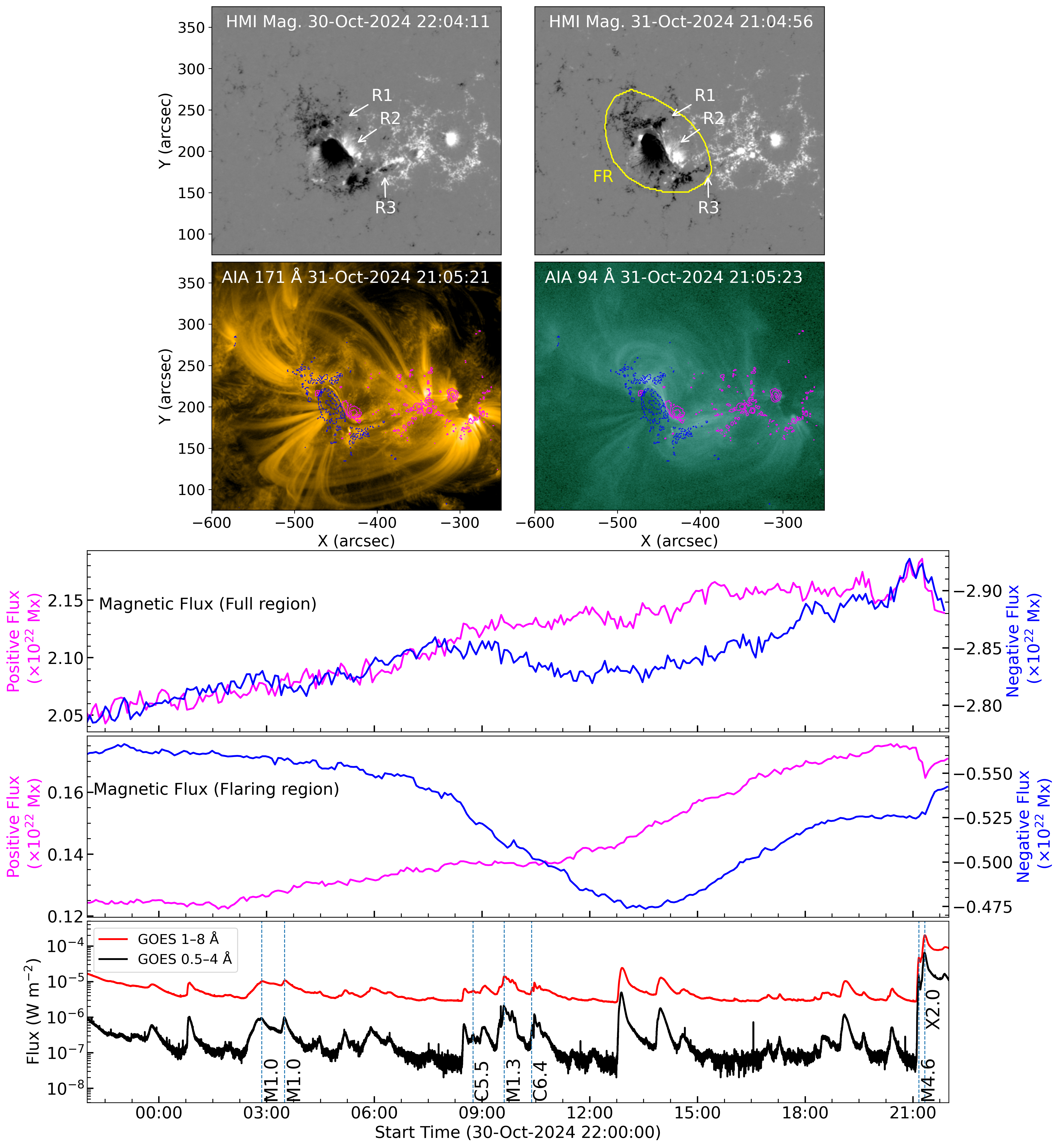}
    \caption{Top panels: SDO/HMI line-of-sight magnetograms on 30 October 2024 (22:04~UT) and 31 October 2024 (21:04~UT), showing regions R1, R2, and R3 that exhibited signiﬁcant changes and the flaring region (FR, yellow contour). Middle panels: Co-aligned AIA 171~\AA\ and 94~\AA\ images overlaid with magnetic field contours ($\pm 500$, $\pm 1000$, $\pm 1500~\mathrm{G}$) on 31 October 2024 (21:04~UT). Bottom panels: temporal evolution of positive (magenta colour) and negative (blue colour) magnetic flux for the full active region (top) and the flaring region (middle), together with GOES 1--8~\AA\ and 0.5--4~\AA\ soft X-ray flux (bottom). Vertical dashed lines mark the times of associated flare events. The animation begins at 22:04 UT on 2024 October 30 and ends at 20:28 UT on 2024 October 31. The real-time duration of the animation is 13 s. The animation highlights the progressive changes in the photospheric magnetic field, including flux emergence and cancellation.\\
    (An animation of this ﬁgure is available in the online version of the article.)}
    \label{fig:magnetic_flux}
\end{figure}

Figure~\ref{fig:magnetic_flux} shows the structure and evolution of the photospheric magnetic field and the associated magnetic flux variations in AR 13878 during the flare. We note that the event occurred during the developing phase of the AR, marked by rapid sunspot area growth from $\approx$80 to $\approx$350 millionths of the solar hemisphere as reported by the Space Weather Prediction Centre \footnote{https://www.swpc.noaa.gov/}. The top panels present SDO/HMI line-of-sight magnetograms taken on 30 October 2024 (22:04 UT) and 31 October 2024 (21:04 UT); i.e., an interval of 24 hours preceding the event under investigation. Three regions of interest (R1, R2, and R3) are marked, where noticeable changes in the magnetic field are observed. The yellow contour outlines the flaring region (FR). A clockwise rotation is seen along the main PIL within R2, while R1 shows an increase in negative polarity and R3 shows an increase in positive polarity.

The middle panels display co-aligned AIA 171~\AA\ and 94~\AA\ images with overlaid photospheric magnetic field contours (positive in magenta and negative in blue) from 31 October 2024 (21:04~UT). These images highlight how the coronal structures are related to the underlying magnetic field distribution. The lower panels show the temporal evolution of the positive (magenta) and negative (blue) magnetic flux for both the full active region and the flaring region. In the full active region, both polarities gradually increase after about 09:00~UT, the negative flux shows a slight decrease, while the positive flux continues to rise. In the flaring region, the positive flux steadily increases, whereas the negative flux first decreases and then begins to rise after 13:00~UT. The bottom panel presents the GOES soft X-ray flux in the 1--8~\AA\ and 0.5--4~\AA\ channels. Vertical dashed lines mark the timings of the flares associated with AR~13878.

\subsubsection{Magnetic configuration and topology}
\label{sec:xrt_aia}

\begin{figure}
    \centering
    \includegraphics[width=0.75\linewidth]{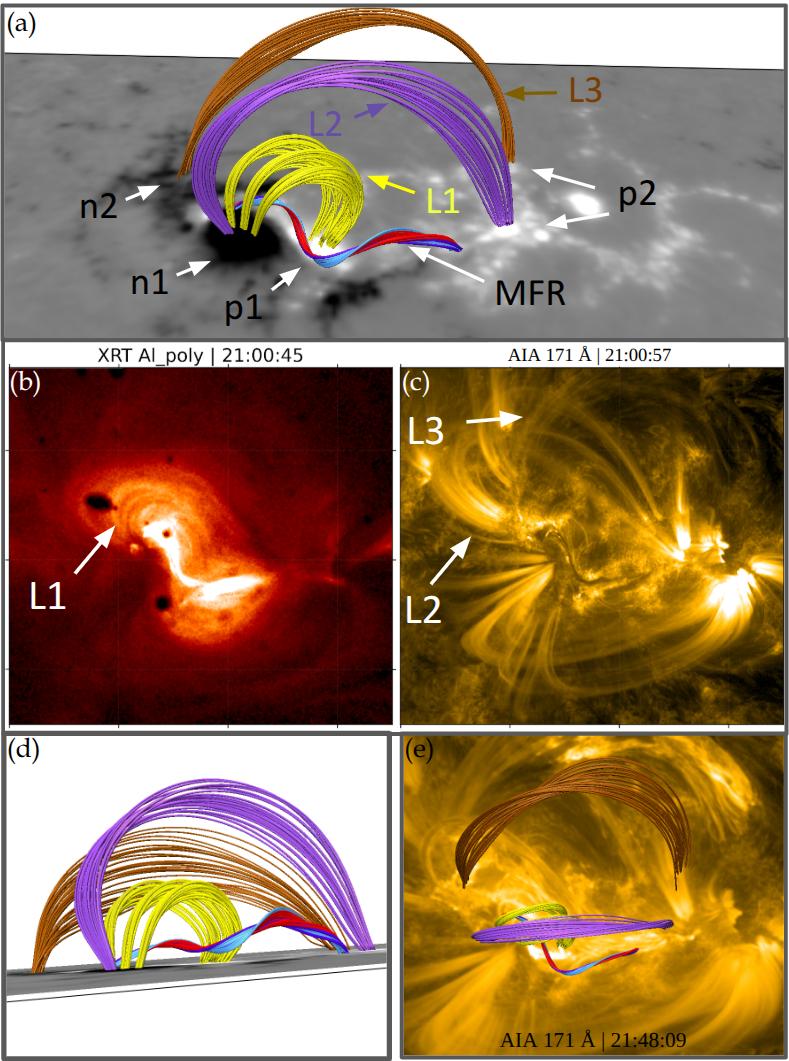}
    \caption{Multi-panel representation of the flaring active region at approximately 21:00~UT during the pre-flare stage. The identified magnetic structures, including the loop systems L1, L2, and L3, as well as the low-lying flux rope (FR), are labelled. Panels (a) show magnetic field lines derived from the NLFFF extrapolation overlaid on the HMI photospheric magnetogram, highlighting the three-dimensional magnetic topology of the region. The yellow, purple, and brown field lines represent the loop systems L1, L2, and L3, respectively, whereas the red-blue field lines denote the flux rope (FR). Panel (b) presents the \textit{Hinode}/XRT Al\_poly image, showing hot coronal plasma associated with the active-region core. Panel (c) displays the \textit{SDO}/AIA 171~\AA\ image, revealing the surrounding coronal loop. Panels (d) and (e) show the NLFFF image from different viewing angles without and with the AIA 171 \AA\ filtergram, demonstrating the close agreement between the reconstructed magnetic configuration and the observed coronal loops.}
    \label{fig:xrt_aia}
\end{figure}

As discussed in the previous sections, the flare exhibits the three energy-released episodes, P1, P2, and P3 along with an EUV late phase. To probe the driving mechanism for episodic energy release and the causal relation between them, it is important to examine the coronal magnetic structure of the AR in detail. In view of this, we present multiple views of the AR using direct imaging and NLFFF extrapolation in Figure~\ref{fig:xrt_aia}. 
In Figure~\ref{fig:xrt_aia}a, we identify various magnetic flux structures that play a crucial role in driving the multi-phase evolution of the X-class flare. The modelled field line clearly shows a twisted, low-lying magnetic structure along the PIL, identified as MFR. The MFR together with the core fields erupts and causes the flare initiation (see Figure~\ref{fig:aia_phase1}; Section~\ref{sec:ph1}). We recognised a low-lying magnetic loop system L1 (yellow field lines in Figure~\ref{fig:xrt_aia}a) that immediately constrains the MFR and terminates at opposite polarity n1 and p1. From the same magnetic polarity region, n1, emerges a high-lying loop system L2 (purple field line) which terminates at a distinct and dispersed positive polarity region at the photosphere, marked as p2. Another high-lying loop system, L3 (brown field line), can also be clearly identified, which joins the distant photosphere region n2 with the p2 region. In panels b and c, we show two representative pre-flare coronal images in X-ray (obtained from XRT Al\_poly; \citealt{xrt}) and EUV (AIA 171 \AA) channels, respectively.
The X-ray images readily exhibit hot plasma emission outlining the core region and the associated low-lying loop arcade, which are identified as MFR and L1 magnetic structures in the NLFFF extrapolation (see panel a). It is likely that the MFR, together with L1 form the core field region. The EUV 171 \AA\ images sensitive to low plasma temperature (log~$T$\,=\,5.8) clearly distinguished two sets of high-lying loop systems enclosing the flaring region, which match with L2 and L3 loops in the NLFFF image. To further elaborate on the three-dimensional magnetic structure of AR, we further show the NLFFF image from different viewing angles without and with the AIA 171 \AA\ filtergram in panels d and e, respectively. The magnetic configuration of the AR, evidenced from various panels of Figure~\ref{fig:xrt_aia}, clearly demonstrates that the activation and eruption of MFR is tightly linked with overlying loop systems L1, L2 and L3. 

The inferred magnetic evolution suggests a progressive involvement of magnetic structures of increasing spatial scale. The first phase of energy-release (P1) is associated with reconnection beneath the MFR. Subsequently, the MFR, together with core fields (L1), undergoes upward activation. The strong XRT brightening indicates that the earliest energy release occurred within this compact core region involving MFR lying along the PIL and overlying L1 core fields. As the flux rope undergoes upward activation, it perturbs the surrounding magnetic environment and subsequently destabilises the neighbouring loop system L2, triggering external reconnection and producing the second energy-release phase (P2). Subsequently, the perturbation propagates to the overlying arcade, L3, where further external reconnection and plasma heating generate the third energy-release phase (P3), and this arcade also forms a side-lobe structure that channels plasma downflows, as observed in Figure~\ref{fig:171_flow}. Therefore, the three episodes can be interpreted as a cascade of magnetic reconnection events, in which the initial activation of the MFR and surrounding core fields triggers successive reconnection in progressively larger magnetic systems (L1 $\rightarrow$ L2 $\rightarrow$ L3) (see Figure~\ref{fig:AIA_GOES_combined_lightcurve} and associated animation).

\subsubsection{Overlying magnetic environment and the MFR}
\label{sec:decay_index}

\begin{figure}
    \centering
    \includegraphics[width=1\linewidth]{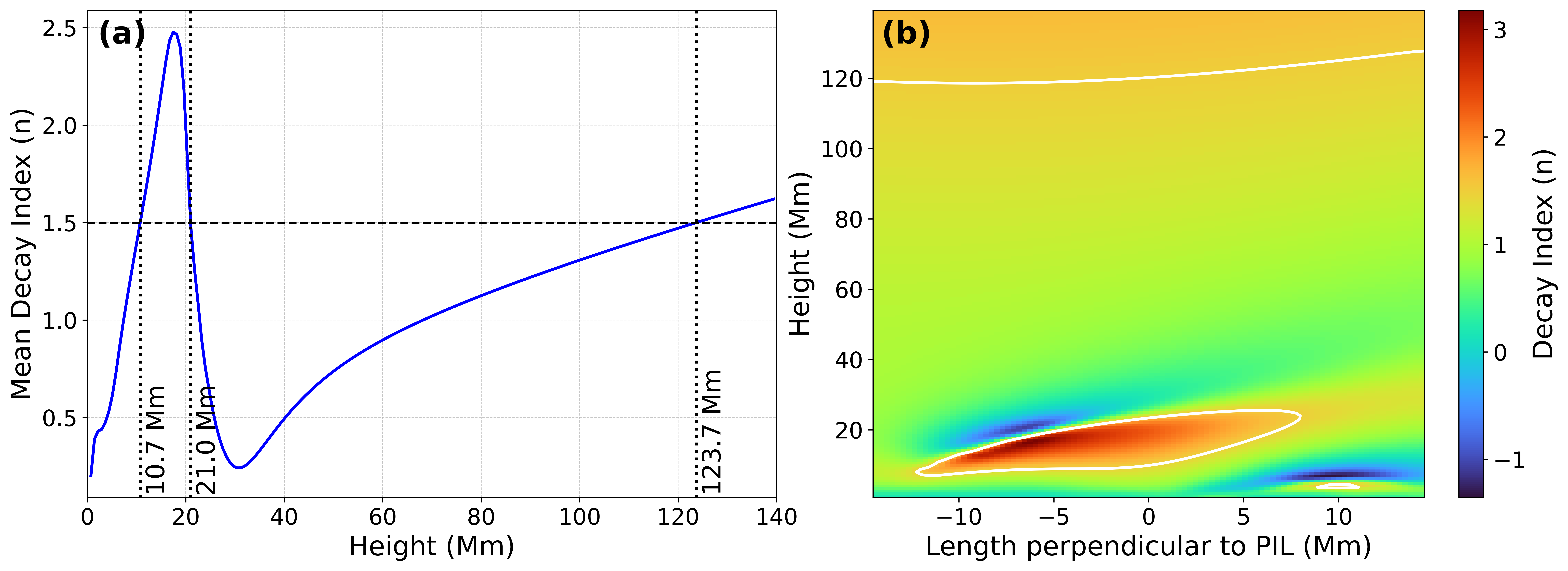}
    \caption{Decay-index analysis of the coronal magnetic field above the PIL. Panel (a): Variation of the mean decay index (\(n\)) with height. The horizontal dashed line denotes the critical torus-instability threshold (\(n=1.5\)), while the vertical dotted lines indicate the heights at which the decay index crosses the critical threshold. Panel (b): Two-dimensional distribution of the decay index in a vertical plane perpendicular to the PIL. The white contour corresponds to \(n=1.5\), separating torus-stable and torus-unstable regions.}
    \label{fig:decay_index}
\end{figure}

The magnetic decay index quantifies how rapidly the external magnetic field decreases with height and is a key parameter governing the onset of torus instability. We computed the decay index ($n$) within the extrapolation volume. The decay index is defined as
\begin{equation}
n = - \frac{d \log B_{\mathrm{h}}}{d \log h},
\end{equation}
where $B_{\mathrm{h}} = \sqrt{B_{x}^{2} + B_{y}^{2}}$ represents the horizontal component of the magnetic field, and $h$ denotes the height above the photosphere. To evaluate the spatial variation of the decay index, we first identified the PIL in the photospheric magnetogram. The decay index was then calculated both along the PIL and in the direction perpendicular to it. We find similar variation in decay index both along and perpendicular to the PIL. Therefore, in Figure~\ref{fig:decay_index} we show the variation of the decay index in the perpendicular plane of the PIL only. This approach allows us to characterise the vertical and lateral variation of the overlying magnetic field and to assess whether the conditions necessary for torus instability are satisfied in the vicinity of the flux rope.

Figure~\ref{fig:decay_index} shows the variation of the decay index as a function of height above the PIL, along with its two-dimensional spatial distribution in a vertical plane perpendicular to the PIL. The Figure~\ref{fig:decay_index}a presents the decay index profile averaged along the direction perpendicular to the PIL. The dashed horizontal line marks the critical threshold for torus instability ($n \approx 1.5$; \cite{Kliem2006}). The spatial distribution of the decay index shown in Figure~\ref{fig:decay_index}b reveals a complex layered structure above the flux rope axis. At very low heights, an extended region of enhanced decay index is observed, highlighted by the red patch and enclosed by the white contour. This high-$n$ region is immediately surrounded by a layer where the decay index decreases to very low values ($n \approx -1$), as indicated by the blue region. At greater heights, the decay index gradually increases with altitude and eventually reaches the critical value of $n \approx 1.5$ only in the higher corona, as marked by the upper white contour. 
From the NLFFF extrapolation, we estimated the approximate heights of the flux rope (FR) and the overlying loop systems L1, L2, and L3. The flux rope is a low-lying structure with a height of $\approx$ 6-10~Mm in the core region. The compact loop system L1 extends to a height of $\approx$ 20~Mm, while the larger overlying loop systems L2 and L3 attain heights of $\approx$ 48--50~Mm. 

The decay-index analysis provides important insight into the eruptive behaviour of the event. The decay index remains below the critical torus-instability threshold (n=1.5) below a height of approximately 10~Mm, indicating a torus-stable regime. Between heights of about 10 and 21~Mm, the decay index exceeds the critical threshold, corresponding to a torus-unstable region. Since loop system L1 is located within this height range, the upward expansion of the flux rope can perturb and destabilise L1, leading to the first energy-release phase (P1). Above $\sim$21~Mm, the decay index decreases again and forms a saddle-like profile, returning to a torus-stable regime. The larger loop systems L2 and L3 are situated within this stable region, where the overlying magnetic field remains sufficiently strong to suppress further eruption. Consequently, although the flux rope becomes activated and causes the overlying core fields (L1) to expand, it is unable to destabilise the higher overlying arcade systems. The eruption, therefore, fails to satisfy the conditions for sustained torus instability during the P2 and P3 phases. 

The complexity of an MFR can be measured with its twist number ($T_{w}$), to examine whether kink instability could have acted as a trigger for the activation of the flux rope. For this purpose, we computed the magnetic twist number within the NLFFF extrapolation volume. The twist number ($T_{w}$), which quantifies the winding of magnetic field lines around the axis of a flux rope, is defined as \citep{Berger2006}
\begin{equation}
T_{w} = \frac{1}{4\pi} \int_{L} \frac{(\nabla \times \mathbf{B}) \cdot \mathbf{B}}{B^{2}} \, dl,
\end{equation}
where $L$ represents the length of the flux rope. The twist number was evaluated using the IDL-based implementation described by \citet{Liu2016}. The average twist number for the flux rope structure is approximately $T_{w} \approx 1.5$.

\section{Discussion}
\label{sec:discussion}

This study provides a comprehensive multi-wavelength investigation of a confined X2.0 solar flare exhibiting two phases of HXR emission, P1 and P2, followed by EUV late phase along with measurable HXR peak P3 (see Figure \ref{fig:GOES_Xray_Flux_crop1}, \ref{fig:combined1} and \ref{fig:AIA_GOES_combined_lightcurve}). This work is motivated by the novel observations of the solar X-ray emission at high spectral and temporal resolutions obtained from SoLEXS and HEL1OS payloads onboard Aditya-L1. A comprehensive synthesis of multi-instrument data from Aditya-L1 and SDO enables detailed probing of the thermal, non-thermal, magnetic, and dynamical characteristics of the event. In Table~\ref{tab:event_summary}, we discuss the key observational characteristics of the event during the three phases. In the following, we summarise the main scientific results of this work and discuss them in the context of our contemporary understanding of the physics of solar eruptions.

The SoLEXS and HEL1OS light curves reveal two temporally connected phases of HXR emission (see blue and green shaded regions in Figure \ref{fig:combined1}). We note that, while the separation between peak intensity of the two phases is $\approx$10 minutes for GOES 1--8 \AA\ and low-energy (up to 10 keV) light curve from SoLEXS (c.f. see Figure \ref{fig:GOES_Xray_Flux_crop1} and Figure~\ref{fig:combined1}a), with increasing X-ray intensities, the time separation decreases. Notably, a 40-60 keV light curve constructed from HEL1OS/CZT observations (Figure 2c) reveals a separation between the HXR emission peaks to be $\approx$8 minutes. The systematic decrease between the peaks of the two phases of the X-ray emissions essentially points towards the underlying time delay between non-thermal and thermal phenomena. In general, during both phases, impulsive non-thermal emission clearly precedes the gradual thermal counterparts (c.f. panels a and c in Figure~\ref{fig:combined1}). The bulk of hard X-ray emission at energies above 25~keV is commonly used to mark the onset of the impulsive phase of a solar flare, as it indicates the acceleration of electrons due to non-thermal bremsstrahlung \citep[see][]{Ramaty2000}. The first phase (P1) grows up to M4.6 level in the GOES scale and peaks at 21:10 UT. A close examination of the multiband X-ray light curve (see Figure~\ref{fig:combined1}) reveals this phase to be of shorter time-scale ($\sim$6 minutes) yet presents impulsive HXR flux variation and substructure within it. HEL1OS/CZT measurement readily confirms statistically significant HXR flux in the 40-60 keV energy band (Figure~\ref{fig:combined1}c). After a brief decay of flux for about 2 minutes in the X-ray channels -- still remaining at a much higher level than the background and clearly identified in the GOES (Figure \ref{fig:GOES_Xray_Flux_crop1}) and SoLEX/Aditya-L1 (Figure~\ref{fig:combined1}a) light curves -- the flux further builds up from 21:12 UT, indicating the onset of the second phase of the flare. Aditya-L1 X-ray observation clearly reveals that the second phase sustains for a longer period ($\sim$15 minutes) with strong HXR emission up to $\geq$80 keV energies. The GOES flux builds up to X2.0 level and subsequently shows a gradual decline (Figure \ref{fig:GOES_Xray_Flux_crop1}); the SoLEXS observation at low energy (up to 10 keV) closely mimics the GOES 1--8 \AA\ profile (c.f. Figure \ref{fig:GOES_Xray_Flux_crop1} and Figure~\ref{fig:combined1}a). The hardest X-ray band (80-150 keV) peaks at $\sim$21:16 UT, $\sim$4 minutes before the SXR peak P2 at 21:20 UT, identified by GOES and SoLEXS. The presence of two broad and intense HXR bursts during the two temporally connected phases of X-ray emissions implies the occurrence of two large-scale events of magnetic reconnection. While the evolution of the two phases is not a general characteristic of large flares, it is not uncommon as well \citep{Joshi2009, Joshi2018, sMitra2020}.

The validation of the confined nature of this two-phase X-class flare is unambiguously confirmed using coordinated multi-satellite observations. White-light coronagraph data from SOHO/LASCO C2 and C3, with a cadence of 12 minutes, together with higher-cadence 5-minute observations from STEREO-A/COR1 and COR2, reveal no signatures of a coronal mass ejection associated with the event. In addition, an examination of radio dynamic spectra from the WIND/WAVES instruments, SWAVE/STEREO-A, and the ground-based e-CALLISTO spectrometers shows no evidence of Type~II or Type~III radio bursts. The absence of these radio signatures indicates that neither large-scale shock formation nor the opening of magnetic field lines into the interplanetary medium occurred, presenting a general consistency with the confined nature of the flare.

A central feature of this event is the activation of an MFR that ultimately fails to erupt. AIA 94~\AA\ and 131~\AA\ observations reveal the formation and expansion of a hot-channel structure (see panels a1--a5 and b1--b5 in Figure ~\ref{fig:aia_phase1}), while height--time measurements indicate an accelerating rise reaching velocities of $\sim$204~km~s$^{-1}$ and accelerations of a few km~s$^{-2}$ (Figure ~\ref{fig:94_slice_pair}). Thus, the kinematic evolution of MFR suggests that despite the structure experiencing an overall strong upward Lorentz force, it attains only limited heights and does not escape. The overlying coronal arcade, traced by expanding yet closed AIA 171~\AA\ loops (see panels c1--c5 in Figure ~\ref{fig:aia_phase1}), provides strong magnetic confinement that suppresses the eruption. The asymmetric expansion of the MFR (see Figure~\ref{fig:94_slice_pair}) further suggests spatial variations in the constraining overlying magnetic field strength, while the brief acceleration phase followed by stalled motion confirms a reconnection-driven but magnetically confined eruption. In a comprehensive study of a failed eruption, \citet{Kushwaha2015} detected strong EUV emission with multiple MW sources from the extended coronal region during the confinement of the erupting MFR. Furthermore, recent observations indicate that confined flux ropes may continue to evolve after their ascent is halted. \citet{Zheng2023} demonstrated that a confined MFR can interact with underlying sheared arcades through external magnetic reconnection during its downward motion, leading to flux-rope deformation and additional energy release. Such a post-confinement evolution provides a plausible mechanism for sustained coronal heating in confined X-class flares such as the one studied here.
\citet{Zhong2021} found that solar eruptions can fail not only because the overlying field is too strong, but also because the flux rope’s own non-axisymmetric magnetic structure produces a downward Lorentz force that halts its rise. \citet{Wang2007} performed a systematic analysis of X-class flares spanning both eruptive and confined categories and found that confined flares preferentially originate near the magnetic centres of active regions, whereas eruptive flares are more likely to occur toward their edge. Contextually, the present event occurred in a complex $\beta\gamma\delta$ magnetic configuration of the AR with the flare-associated PIL located beneath a strong overlying flux system, close to the central part of the AR (see Figures~\ref{fig:aia_phase1} and \ref{fig:magnetic_flux}).
It is also noteworthy that this large confined eruption occurred in the growing phase of the active region with strong emerging flux regions. As an aside, we mention the statistical results by \citet{Li2024} that confined and eruptive flares occur in comparable numbers during the developing phase of active regions, whereas confined flares dominate during decay.

Apart from the direct observational signatures of confinement, our magnetic-field modelling provides quantitative evidence for the physical mechanism responsible for the failed eruption. The NLFFF extrapolation reveals a low-lying MFR embedded beneath multiple overlying arcade systems (L1--L3; Figure~\ref{fig:xrt_aia}), while the decay-index analysis shows the saddle-like profile and the flux rope encounters only a narrow torus-unstable layer ($n \approx 1.5$) between heights of $\sim$10 and 21~Mm before entering an extended torus-stable regime (Figure~\ref{fig:decay_index}). Previous studies have demonstrated that a saddle-like decay-index profile favours failed eruptions because the overlying toroidal strapping force, after initially weakening with height, strengthens again at larger heights, thereby restoring magnetic confinement and preventing the continued rise of the flux rope \citep{Wang2017,Filippov2020b, Mitra2022}. Consequently, although the MFR undergoes rapid acceleration and triggers magnetic reconnection, it fails to attain the critical height required for sustained torus instability. Instead, the strong overlying arcade suppresses its further ascent, leading to a confined eruption. Such behaviour is consistent with theoretical and observational studies showing that strong overlying magnetic fields can effectively halt erupting flux ropes before they escape into the heliosphere \citep{Torok2004,Torok2005,Sun2015,Luo2022}.

An important aspect of the observations is the initial activation and accelerated rise of the flux rope (Figure~\ref{fig:94_slice_pair}), which was subsequently halted by the strong strapping force exerted by the overlying magnetic field. Therefore, understanding the early eruption phase also warrants attention. A possible clue to the triggering of the initial rise of the flux rope comes from its magnetic twist, which is estimated to be $T_{\mathrm{w}}\approx1.5$, a marginal value for the onset of kink instability (KI). Numerical simulations have shown that the critical twist required for KI depends on the magnetic flux rope geometry, with reported threshold values ranging from approximately 1.5 to 1.75 turns \citep{Fan2003,Torok2004,Guo2010}. Furthermore, although statistical studies suggest that a magnetic twist exceeding approximately 2 turns provides a sufficient condition for KI, many eruptive events exhibit twists between 1 and 2 turns \citep{Duan2019}. Therefore, our measured twist ($T_{\mathrm{w}}\approx1.5$) represents a marginally kink-unstable configuration and suggests that KI may have contributed to the initial activation of the flux rope. However, during the activation of the MFR, magnetic reconnection within the sheared core field, as indicated by early localized brightenings in the AIA 304 Å images (see associated animation for Figure~\ref{fig:AIA_GOES_combined_lightcurve} and Figure~\ref{fig:aia_phase1}(d1)), may also have contributed to its destabilization, suggesting the occurrence of tether-cutting-like reconnection \citep{Moore1992, Moore2001, Joshi2017, Sahu2023, binal2024}.

Magnetic flux ropes are a fundamental element of solar eruption models, and their presence in the low corona has been established by numerous EUV observations, particularly in the hot channels of AIA \citep{Cheng2011, Zhang2012, cheng2013, Joshi2017, Mitra2018, Aaron2019, Mitra2019}. In the present event, expansion of EUV hot channel appears (AIA 94~\AA\ and 131~\AA\ channels) approximately five minutes after the first emission peak P1 (Figure~\ref{fig:aia_phase1}), indicating sustained build-up of structure with the onset and progression of coronal magnetic reconnection. The EUV imaging observations coupled with X-ray temporal signatures shown, in the present study, are consistent with the detection of HXR coronal channels observed in energies up to 25 keV during the active pre-flare phase of an eruptive long-duration flare investigated by \citet{Suraj2020}. Besides hot EUV channels, MFRs are observed to exhibit multi-thermal and multi-component morphologies in the corona, e.g. sigmoid \citep{Rust1996, McKenzie2008, Hema2021}, hot plasma blobs or plasmoids \citep{Shibata1998, Kumar2013, Joshi2016}. A comparison of the hot channels height time profile with associated X-ray emission contains crucial evidence about the linkages between magnetic reconnection and MFR dynametics--the buildup and activation of flux rope during the first phase is caused by the magnetic reconnection, while, during the second phase, it is the flux rope dynamics that mostly controls the magnetic reconnection.

The joint X-ray spectroscopic analysis using SoLEXS and HEL1OS (Figure~\ref{fig:aditya_joint} and Figure~\ref{fig:combined_parameter}) provides additional important insights on thermal and non-thermal process energy released associated with two-phase evolution of the confined eruption. During the first phase (P1), we find a maximum plasma temperature of $\sim$19~MK, which correlates with peak X-ray emission, while the EM continues to rise during this phase, implying a gradual change from localised to widespread plasma heating (Figure~\ref{fig:combined_parameter}). The spectral fitting reveals a soft photon spectrum ($\gamma \approx 4.34$), with detectable emission only up to $\sim$60~keV above the background level, suggesting relatively weak non-thermal electron acceleration. The break energy during this phase undergoes a limited variation ($\sim$11.5~keV and $\sim$14.5~keV). In contrast, the main X2.0 phase (P2) exhibits a significantly higher maximum temperature of $\sim$22~MK, a steep rise in EM and strong hard X-ray emission extending up to $\sim$150~keV above the background. This phase is characterised by a much harder photon spectrum ($\gamma \approx 3.96$), indicating more efficient acceleration of high-energy electrons. A comprehensive study by \citet{Caspi2015} demonstrated that superhot plasma emission remains cospatial with the non-thermal source for a finite duration, implying a close physical connection between the superhot ($\sim$30MK) component and coronal non-thermal electron acceleration, consistent with our observations (see Figure~\ref{fig:combined_parameter}). Furthermore, the increasing trend in low energy cutoff ( $\sim$12~keV to $\sim$16~keV) throughout this phase reflects progressive dominance of thermal emission beyond peak HXR emission, which essentially implies conversion of energy of accelerated electron to the heating of ambient plasma.

Identification of a pronounced and extended EUV late-phase emission (21:50--22:20~UT; Figure~\ref{fig:AIA_GOES_combined_lightcurve}; Table~\ref{tab:event_summary}) is an important characteristic of the event that requires further attention. This phase is characterized by delayed intensity peaks in the AIA 131~\AA\ ($\sim$22:00~UT) and 94~\AA\ ($\sim$22:08~UT) channels. It is noteworthy that the early part of the EUV late phase coincides with the third phase of X-ray emission (P3), peaking at $\sim$21:52~UT, as observed in the GOES and X-ray/Aditya-L1 light curves (see Figure~\ref{fig:GOES_Xray_Flux_crop1}, \ref{fig:combined1} and \ref{fig:AIA_GOES_combined_lightcurve}). The joint SoLEXS--HEL1OS spectral analysis shows that P3 is dominated by weak non-thermal emission, with X-ray emission confined below $\sim$30~keV and a relatively soft photon spectrum ($\gamma \approx$ 7.3--9.5). The presence of non-thermal X-rays during P3 implies a reconnection-induced episode of particle acceleration and heating. The X-ray spectroscopic evidence of particle acceleration, together with the temporal separation between P3 and the EUV late-phase peaks, indicates that they represent different stages of the post-flare energy evolution. We emphasise that the EUV emission in the AIA 131~\AA\ and 94~\AA\ channels continues to increase after the X-ray energy episode P3 and eventually exceeds the peak flux in these EUV channels during the decay phase of the X-class flare. Investigating the origin of the EUV late phase remains one of the important topics in solar physics. A study based on a large sample of flares using 15 years of SDO observations suggests that the EUV late phase is not uncommon, yet it is not a general feature of flares, with an occurrence rate of approximately 8\% \citep[e.g.,][]{Greatorex2025}.

The EUV late phase was first identified by \citet{Woods2011} as a delayed secondary enhancement in warm coronal EUV emission originating from a second system of longer flare loops. Two principal mechanisms have subsequently been proposed to explain this phenomenon. The first attributes the delayed EUV emission to the prolonged cooling of longer flare loops that are heated during the main energy-release phase \citep{Dai2018,Li2024a}. The second invokes additional heating during the decay phase through continued magnetic reconnection in the overlying magnetic system \citep{Sun2013,Zhou2019a}. In the present event, our results suggest that both mechanisms are likely to contribute. During the X-class flare, we found that the MFR underwent a failed eruption due to confinement by a strong overlying magnetic arcades. This is further supported by the presence of multiple lanes of bright downflows originating within the large coronal loop system, with projected velocities ranging from approximately 100 to more than 500~km~s$^{-1}$ (Figures~\ref{fig:171_flow}), indicating continued dynamical evolution of the outer flaring corona during the extended decay phase. It is likely that the released magnetic energy was deposited in the flaring corona, within the large-scale coronal loops, rather than being carried away by a CME. These longer loops are expected to cool more slowly, naturally producing delayed emission in the temperature response of the AIA 131~\AA\ and 94~\AA\ channels, consistent with the long-cooling scenario proposed by \citet{Dai2018} and \citet{Li2024a}. However, the occurrence of a subtle yet clear non-thermal X-ray enhancement during P3, preceding the EUV late phase, implies a reconnection-induced episode of particle acceleration and heating. In both scenarios proposed for the origin of the EUV late phase, the secondary heating is expected to occur in the extended flaring regions, away from the core fields. The spatial connection of the energy-release site during the EUV late phase with respect to the core flaring region is clearly evident from the spatially resolved AIA light curves (Figures~\ref{fig:box_aia}), which show plasma heating at greater heights and in larger coronal loops.

For a complete understanding of the confined eruption, particularly the forceful halt of the ascending flux rope by the overlying field and its subsequent deformation, we examine in AIA 304 \AA\ images to investigate the chromospheric response to energy released and the dynamic process occurring in the corona. The AIA 304 \AA\ images clearly reveal the reappearance of a dark filament along the PIL, which undergoes impressive dynamical evolution (see Figure~\ref{fig:aia_phase2} and AIA animation between 21:25 UT to 21:50 UT associated with Figure~\ref{fig:AIA_GOES_combined_lightcurve}) -- a feature commonly identified as filament reformation. This feature is commonly observed in partial or failed eruptions, where plasma falls back following the flux rope confinement
 \citep{Gilbert2007, JoshiN2014, Cheng2018}. However, as in the present study, the detailed reformation process has rarely been captured. In our observations, it is noteworthy that following the confinement of the MFR, another filament reformed at the same location as the confined one (Figure~\ref{fig:aia_phase1} and Figure~\ref{fig:aia_phase2}).

\begin{table*}[!hbt]
\centering
\caption{Chronological summary of the confined X2.0 flare on 2024 October 31.}
\label{tab:event_summary}
\footnotesize
\renewcommand{\arraystretch}{1.2}

\begin{tabular}{c c c c l}
\hline
\textbf{Sr. No.} &
\textbf{Phases} &
\textbf{Time Interval} &
\textbf{Dominant Emission} &
\textbf{Key Observations} \\
\hline

1 & P1 & 21:06-21:12~UT & X-ray and (E)UV &
\parbox[t]{5.5cm}{X-ray emission up to $\approx$ 60 keV with  T $\approx$ 19 MK and $\gamma \approx 4.34$; intense EUV core with hot channel (MFR) and bright overlying loops.} \\

2 & P2 & 21:12-21:27~UT & X-ray and (E)UV &
\parbox[t]{5.5cm}{HXR emission extends to $\approx$150 keV with T $\approx$ 22 MK and $\gamma \approx 3.96$; accelerating rise of MFR--hot channel. } \\

3 & Extended decay phase & 21:27-22:40~UT & EUV with brief X-ray enhancement &
\parbox[t]{5.5cm}{Halt and confinement of the rising hot channel MFR by overlying loops; brief X-ray enhancement (P3; summarised in 3a) and build-up of strong EUV emission (EUV late phase; summarised in 3b). } \\

\hline

3a & P3 & 21:49-21:54~UT & X-ray &
\parbox[t]{5.5cm}{Subtle enhancement in X-ray emission up to $\approx$ 30 keV with T $\approx$ 15 MK and $\gamma \approx 7.3$. } \\

3b & EUV late phase & 21:50-22:20~UT & EUV (94~\AA\ and 131~\AA\ channels) &
\parbox[t]{5.5cm}{Intensity peaks in the AIA 131~\AA\ ($\sim$22:00~UT) and 94~\AA\ ($\sim$22:08~UT) channels; emission from extended coronal region, away from core, and large overlying loops.} \\

\hline
\end{tabular}
\end{table*}

\section{Summary and Prospects}
\label{sec:conclusions}

The paper uniquely describes the process involved in a long-duration confined flare exhibiting two broad phases of HXR emission. The study demonstrates that confined flares can exhibit eruptive-like signatures without producing a CME. 
 The joint spectroscopic analysis using SoLEXS and HEL1OS onboard Aditya-L1 reveals that the first impulsive phase attains a maximum temperature of $\sim$19 MK with a subtle yet distinct non-thermal characteristic ($\gamma \approx 4.34$) and detectable emission up to $\sim$60 keV, whereas the second phase reaches $\sim$22 MK and exhibits a much harder spectrum ($\gamma \approx 3.96$) extending to $\sim$150 keV.  
EUV imaging from AIA/SDO reveals the activation of a hot-channel magnetic flux rope approximately five minutes after the first HXR peak, which accelerates to velocities of $\sim$204 km s$^{-1}$ but halts to expand beyond the overlying large-scale magnetic loops of the AR. During the gradual phase, a distinct third enhancement is identified as weak non-thermal emission confined below $\sim$30 keV, accompanied by a significant widespread enhancement of EUV emission in the hot channels (AIA 94~\AA~and 131~\AA). The detection of asymmetric flux-rope expansion, multiple downflows with speeds of $\sim$100–500 km s$^{-1}$, and filament reformation along the same PIL highlights the role of post-confinement magnetic restructuring in sustaining long-duration coronal heating within a closed magnetic system. The study clearly demonstrates dual-stage particle acceleration, %extended late-phase heating
EUV late phase, hot-channel activation, and filament reformation---all commonly associated with eruptive flares---can occur in a fully confined magnetic environment. 
The identification of an extended EUV late phase is another highlight of the event, in which we find evidence for both heating by magnetic reconnection and subsequent slow cooling due to plasma confinement by the large overlying loops. The interaction of the erupting MFR with the large overlying loops also gives rise to multiple lanes of bright downflows, thereby providing another source of sustained heating during the prolonged decay phase.

From an instrumental perspective, this study showcases the powerful complementarity between solar X-ray observations from Aditya-L1 and (E)UV imaging from SDO/AIA. The joint spectroscopy from SoLEXS and HEL1OS provides high-cadence, broad-band coverage, enabling us to disentangle thermal and non-thermal contributions and to track the evolution of key spectral parameters. When combined with multi-thermal EUV imaging and photospheric magnetic data, this capability opens a new window on the detailed physics of both confined and eruptive flares. Future studies using a larger sample of Aditya-L1 events will enable us to place this flare in a broader statistical context, quantify the frequency of such dual-episode, confined X-class flares, and investigate how the interplay between acceleration and heating depends on the underlying magnetic topology and coronal environment.

\begin{acknowledgments}
We sincerely thank the anonymous referee for the careful reading of the manuscript and for the constructive and insightful comments, which have significantly improved the quality and clarity of this work. We thank the Aditya-L1 and the Solar Dynamic Observatory (SDO) for their open data policies. Aditya-L1 is an observatory-class mission, which is fully funded and operated by the Indian Space Research Organisation (ISRO). We acknowledge the use of data from the Aditya-L1 mission, archived at the Indian Space Science Data Centre (ISSDC), accessible via \url{https://pradan.issdc.gov.in/al1/}. The SDO is NASA’s mission under the Living With a Star (LWS) program. Research at the Physical Research Laboratory (PRL) is supported by the Department of Space, Government of India.

\end{acknowledgments}

%% For this sample we use BibTeX plus aasjournalv7.bst to generate the
%% the bibliography. The sample7.bib file was populated from ADS. To
%% get the citations to show in the compiled file do the following:
%%
%% pdflatex sample7.tex
%% bibtext sample7
%% pdflatex sample7.tex
%% pdflatex sample7.tex

\bibliography{sample701}{}
\bibliographystyle{aasjournalv7}

%% This command is needed to show the entire author+affiliation list when
%% the collaboration and author truncation commands are used.  It has to
%% go at the end of the manuscript.
%\allauthors

%% Include this line if you are using the \added, \replaced, \deleted
%% commands to see a summary list of all changes at the end of the article.
%\listofchanges

\end{document}